\documentclass[onecolumn,usenatbib]{mn2e}

\usepackage{graphicx}
\usepackage{dcolumn}
\usepackage{amsmath}
\usepackage{amssymb}
\usepackage{natbib}

\newif\ifAMStwofonts

\usepackage{amssymb}
\usepackage{epstopdf}
\usepackage{bm}
\def\simlt{\lower.5ex\hbox{$\; \buildrel < \over \sim \;$}}
\def\simgt{\lower.5ex\hbox{$\; \buildrel > \over \sim \;$}}
\def\l#1{\left#1}
\def\r#1{\right#1}

\def\bd#1{\bm{#1}}

\def\const{\hbox{{\sc const}}}

\def\ds{\displaystyle}

\newcommand{\etal}{{et~al.}}
\def\aa{{A\&A}}

\def\apj{{ApJ}}
\def\apjl{{ApJ}}
\def\apjs{ ApJS}

\def\mnras{MNRAS}

\def\pr{Phys.\ Rep.}
\def\prd{PRD}

\title[Maximum likelihood algorithm for parametric component separation]{
Maximum Likelihood algorithm for parametric component separation in CMB experiments}

\author[Radek Stompor, Samuel Leach, Federico Stivoli, Carlo Baccigalupi]
{Radek Stompor$^{1}$, Samuel Leach$^{2}$, Federico Stivoli$^{2}$, 
Carlo Baccigalupi$^{2}$\\
$^{1}$ CNRS, UMR 7164, Laboratoire Astroparticule \& Cosmologie, 10 rue A. Domon et L. Duquet,
F-75205 Paris Cedex 13, France
 \\
$^{2}$ SISSA/ISAS, Astrophysics Sector, Via Beirut, 4, 
and INFN, Sezione di Trieste, Via Valerio 2, I-34014 Trieste, Italy}

\begin{document}

\maketitle

\date{\today}    

\begin{abstract}
We discuss an approach to the component separation of microwave,
multi-frequency sky maps as those typically produced from Cosmic
Microwave Background (CMB) Anisotropy data sets.  The algorithm is
based on the two step, parametric, likelihood-based technique recently
elaborated on by \citet{eriksen_etal_2006}, where the foreground
spectral parameters are estimated prior to the actual separation of
the components.  In contrast with the previous approaches,
we accomplish the former task with help of an analytically-derived
likelihood function for the spectral parameters,
which, we show, yields estimates equal to the maximum likelihood
values of the full multi-dimensional data problem.
We then use these estimates to perform the
second step via the standard,
generalised-least-square-like procedure.  We demonstrate that the proposed
approach is equivalent to a direct maximization of the full data
likelihood, which is recast in a computationally tractable form. 
We use the corresponding curvature matrices  to characterise 
statistical properties of the recovered parameters. We incorporate 
in the formalism some of the essential features of the CMB data sets, such as
inhomogeneous pixel domain noise, unknown map offsets as
well as calibration errors and study their consequences for the separation.
We find that the calibration is likely to have a dominant effect on the precision
of the spectral parameter determination for a realistic CMB experiment.
We apply the algorithm to simulated data and discuss the results.
Our focus is on partial-sky, total-intensity and polarization, CMB experiments
such as planned balloon-borne and ground-based efforts, however, the
techniques presented here should be also applicable to the full-sky
data as for instance, those produced by the WMAP satellite and
anticipated from the Planck mission.
\end{abstract}

\maketitle

\section{Introduction and motivation}

\label{sect:introAndMot}

The measurement and characterization of the Cosmic Microwave 
Background (CMB) polarization anisotropies is an important goal of 
the present day observational cosmology. It poses a significant experimental 
challenge which is aimed at by an entire slew  
of currently operating and forthcoming experiments such as 
PolarBeaR, QUIET, and Clover, observing from the ground, 
EBEX and Spider, observing from the balloon platform,
and WMAP and Planck observing from space\footnote{
PolarBeaR: {\tt http://bolo.berkeley.edu/polarbear/science/science.html},
QUIET: {\tt http://quiet.uchicago.edu/},\\
Clover: {\tt http://www-astro.physics.ox.ac.uk/research/expcosmology/groupclover.html},\\
EBEX: {\tt http://groups.physics.umn.edu/cosmology/ebex/},
Spider: {\tt http://www.astro.caltech.edu/$\sim$lgg/spider\_front.htm},\\
WMAP: {\tt http://map.gsfc.nasa.gov/},
Planck: {\tt http://www.esa.int/esaSC/120398\_index\_0\_m.html}.\\
See also {\tt http://lambda.ngfc.nasa.gov} for a complete list of the ongoing and forthcoming CMB experiments.
}.

The sensitivity of modern CMB experiments is rapidly increasing
as the instruments incorporate more detectors and observe the sky for
longer periods of time.  Consequently, instrumental systematics and
foreground emissions have come to the forefront as the major limiting
factor in the effort of detecting and characterizing the primordial,
cosmological signals.  For total intensity measurements, recently 
extended down to arcminute scales by
ACBAR \citep{reichardt_etal_2008}, and at current
levels of sensitivity, the diffuse galactic foregrounds
have turned out luckily to be quite benign.
However diffuse foreground emission remains a serious concern for CMB
polarimetry experiments.

CMB polarization signal can be decomposed into
a gradient `$E$ mode' component, which is expected to be dominated by
adiabatic scalar perturbations, and into a curl `$B$ mode'
contribution, which can be generated by primordial gravitational waves
\citep{kamionkowski_etal_1997,zaldarriaga_seljak_1997} and lensing of
the $E$ mode by intervening structures \citep{zaldarriaga_seljak_1998, lewis_etal_2006}.
Recent experiments targeting low foreground emission regions of sky
are successfully measuring the $E$ mode of CMB on degree and
sub-degree angular scales, most recently by CAPMAP \citep[for a compilation of recent results,
see][]{bischoff_etal_2008}. However on large angular scales WMAP have
reported a foreground contamination exceeding both the $E$ and $B$
modes of the CMB in the frequency range of the minimum Galactic
emission \citep{page_etal_2007}. Recent analyses of the WMAP five-year
data \citep{gold_etal_2008,dunkley_etal_2008} have seen an renewed
focus on studying the impact of foreground subtraction errors on the
determination of cosmological parameters, using a variety of analysis
methods and assumptions about the data.

Multi-frequency component separation methods differ in the extent to which they
combine the data with `external' sources of information, and the way
in which the components are modeled. The approaches studied to date include
methods based on the internal template subtraction \citep[e.g.][]{bennett_etal_1992, hansen_etal_2006},  those exploiting statistical independence of (some of) the sky components
\citep[e.g.][]{delabrouille_etal_2003, maino_etal_2007, bonaldi_etal_2006},  
those invoking the maximum-entropy principle \citep[e.g.][]{stolyarov_etal_2005},
or performing a parametric fit to the data \citep[][]{brandt_etal_1994, eriksen_etal_2006}. 
Those latter methods, called parametric,
are distinguished by the explicit modeling of the frequency scaling of all components
 and typically by the large number of free parameters describing
the component amplitudes. 
See the Planck Working Group 2 work, \citet{leach_etal_2008}, for more information about
 the component separation methods and a comparison of their performance.

A common requirement for any component separation technique for use in
cosmological analyses is the capability of propagating errors due to
foreground subtraction, while having the flexibility to incorporate
foreground modeling and external constraints in a transparent
way. Against the backdrop of these two requirements, we will consider
the parametric approach to the foreground separation problem.

\subsection{Parametric component separation}

The defining premise of the parametric approach is an assumption that
the functional form of the frequency scaling for all relevant
components is known, and our ignorance can be quantified by means of
relatively few, though non-linear, spectral parameters. The parameters of the model
are determined via a fitting procedure, often performed on a
pixel-by-pixel basis.
The attractiveness of this approach lies in its simplicity, the flexibility of the
parametrization schemes and possibility of phrasing the separation
problem in a coherent maximum likelihood form
\citep[][]{jaffe_et_al_2004}. Its strength lies in exploiting
nearly optimally the prior knowledge of the frequency scaling of
different components, which in several instances is quite well known,
while ignoring the information which is less known and more debatable.
Its weakness is related to the difficulties in performing the
non-linear and high-dimensional parameter fits especially in the
realm of the CMB data, which are usually characterised by low
signal-to-noise ratio on the pixel scale, and numerical efficiency,
given a large number of pixels, ${\cal O}(10^7)$, anticipated from the
next generation of the CMB experiments.

In a recent paper, \citet{eriksen_etal_2006} reconsidered a parametric
approach to the component separation task originally introduced by 
\citet{brandt_etal_1994} and proposed its rendition 
which was shown to be both numerically efficient and stable in the application
to the nearly full-sky simulated Planck and WMAP data. In
their method, the numerical efficacy and stability is achieved by
splitting an estimation of the parameters and component maps 
into two separate steps. 
The spectral parameter fitting is performed pixel-by-pixel on the
input data which are first smoothed and underpixelized during a
preprocessing stage. This enhances the signal-to-noise ratio of the data and
thus improves the stability and robustness of the fitting
procedure. It also makes the procedure more numerically tractable as
fewer fits have to be performed. The fitting criterion, employed by
\citet{eriksen_etal_2006}, uses a likelihood function, and involves
the (non-linear) spectral parameters as well as (linear) component
amplitudes of the smoothed input maps. In the modern twist to the
method, they propose to use a Monte-Carlo Markov Chain (MCMC) sampling
technique as a way to determine the best-fitting parameters, and which
elegantly allows to marginalise over low resolution components leaving
one with the estimates of the spectral parameters. The latter are
taken to be the averages rather than peak values of the sampled
likelihood. On the second step of the method, the recovered
low-resolution parameters are then applied to the corresponding pixels
of the full resolution maps to render the high resolution component
map estimates. More details on the method can be found in
\citet{eriksen_etal_2006}.
In this paper, we modify the parametric method of
\citet{eriksen_etal_2006} and devise the genuinely maximum likelihood
parametric algorithm, while preserving the efficiency of its two step
design. We then discuss in some detail the general features of the
technique, and also demonstrate how it lends itself to a number of
generalizations relevant for the analysis of actual observational
data, such as calibration errors and arbitrary offsets of the input
maps. 

The plan of the paper is as follows. In the remaining part of the
Introduction, we define our notation. In
Section~\ref{section:dataSet} we describe the simulated sky data which
we use in the examples discussed throughout the paper. In
Section~\ref{sect:mlBasics} we cast the parametric approach to
component separation as a maximum likelihood problem, showing how a
two step procedure allows to recover frequency scaling and spatial
patterns of the components. In Section~\ref{sect:mlExts} we extend the
formalism by considering the presence of offsets and calibration
errors into the data, and Section~\ref{sect:concl} contains our
conclusions.

\subsection*{Notation}

We use bold typeface to mark vectors and matrices, written in lower and upper
case letters respectively.  A vector corresponding to a pixel, $p$, or a matrix --
to a pair of pixels, $p,q$, is always given a subscript $p$ or $pq$,
respectively.  (However, if $p=q$ in the latter case we usually drop
one of the $p$.) Such an object will usually contain all the values
for all relevant Stokes parameters and frequency channels (or
alternately, sky components) characterizing a chosen pixel on the
sky. Multi-pixel vectors are always composed of pixel-specific
vectors concatenated together so that the information related to a
first pixel precedes that of a second one etc.  The multi-pixel matrices
are made of pixel pairs specific blocks arranged in an analogous
manner.

\section{Simulated data set}

\label{section:dataSet}

\begin{figure}
\begin{center}
\includegraphics[height=1.75in, angle = 0]{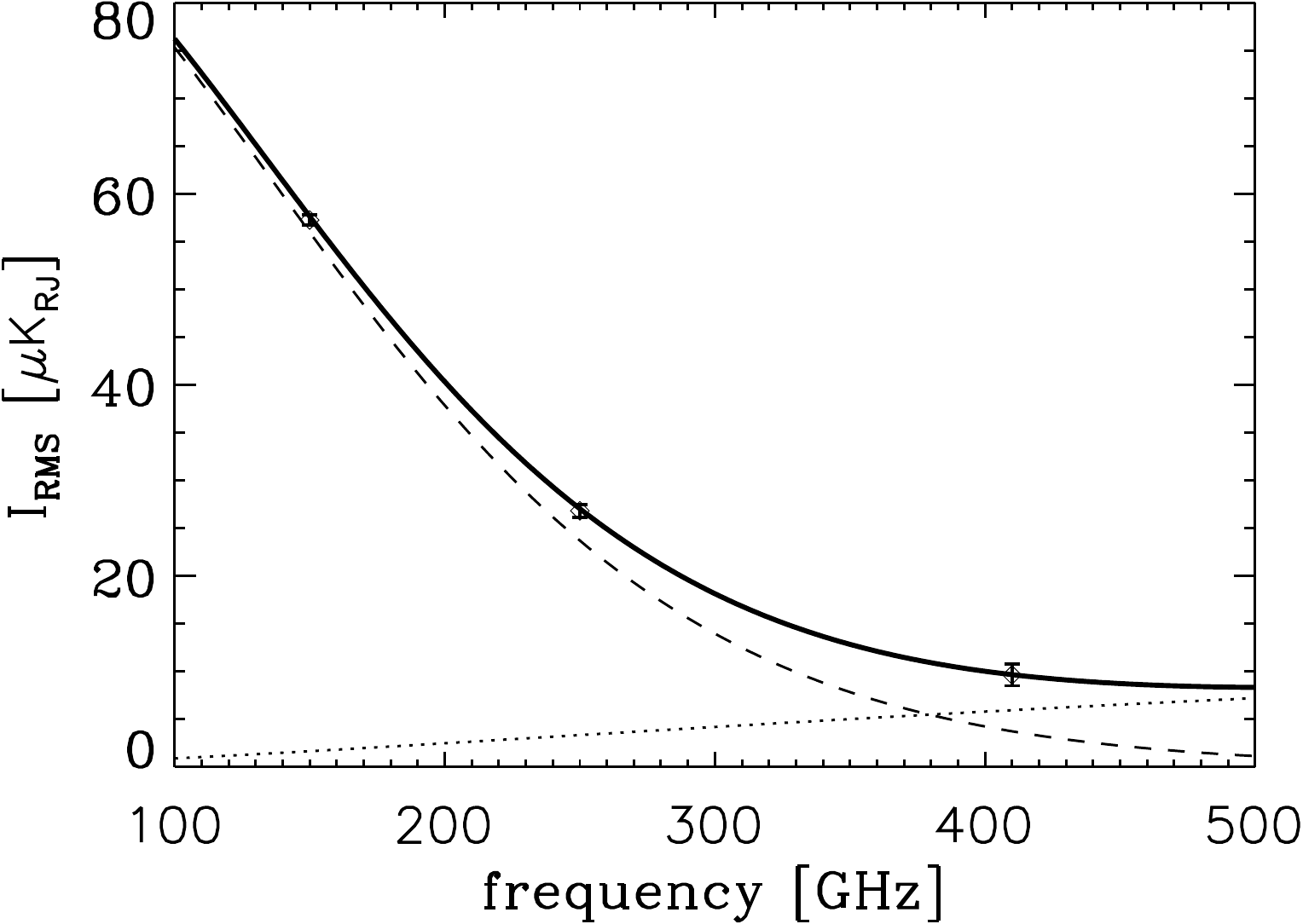}
\includegraphics[height=1.75in, angle = 0]{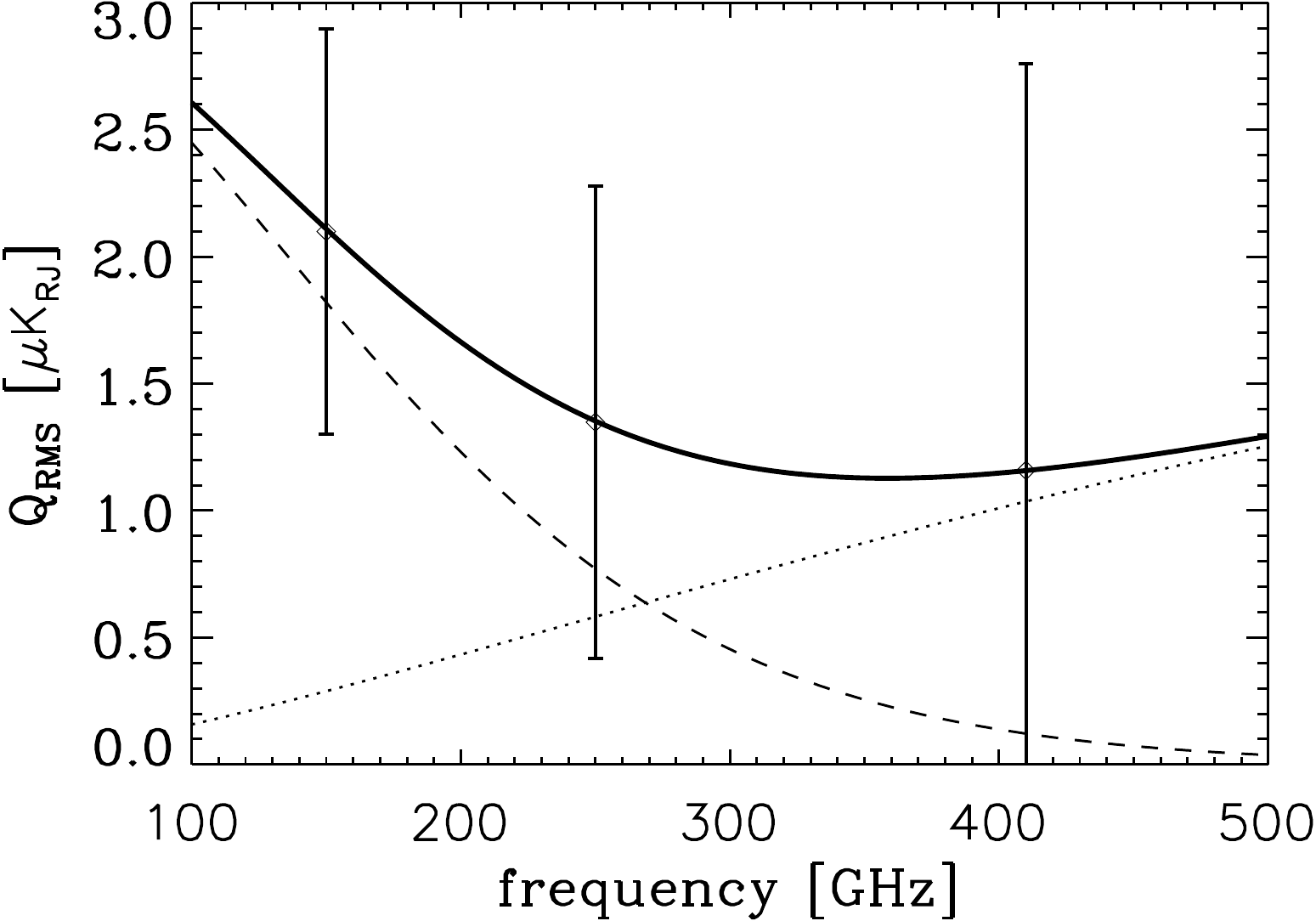}
\caption{The RMS of the simulated signal amplitudes for the CMB (dashed) and 
thermal dust (dotted line) smoothed with an $8'$ beam, for  total intensity $I$ and
Stokes $Q$ parameter ({\bf left and right panels}) as a function of frequency. 
The noise RMS per $3.4'$ pixel ($n_{side} = 1024$) is also shown for the frequency bands
of the experiment we consider here.
\label{fig:TQrms}}
\end{center}
\end{figure}

We simulate the CMB and thermal dust polarized emission
on a patch of sky with an area of 350 square degrees centered at
RA=60$^{\circ}$, Dec=-50$^{\circ}$, adopting the HEALPix\footnote{\tt
http://healpix.jpl.nasa.gov} pixelisation scheme \citep{gorski_etal_2005}
with $3.4'$ pixels ($n_{\rm side}$=1024).
This region of sky is known to be characterised by low dust
emission in total intensity, having been observed by Boomerang
\citep{masi_etal_2006,montroy_etal_2006} and QUaD \citep[][]{ade_etal_2007}.
The polarized emission in this region of sky and in this frequency range is low
but to date has not been as well characterised.
This area is close to the anticipated EBEX
field \citep{oxley_etal_2004}, which we will take as our
reference experiment. Specifically we simulate three
channels at 150, 250 and 410~GHz each with $8'$ resolution, and a noise
level with RMS per pixel of $0.56$, $0.66$, $1.13$ $\mu$K
respectively, in antenna (Rayleigh-Jeans) units; for Stokes $Q$ and $U$ the noise level
is a factor $\sqrt{2}$ higher. The noise is uncorrelated between pixels, frequency
channels and Stokes parameters. The adopted here noise levels correspond to the deep
integration anticipated for $B$ mode CMB experiments. The CMB sky is a
Gaussian realisation of the \citet{spergel_etal_2007}
best-fitting $C_{\ell}$, except with tensor perturbations added in at a
level $T/S=0.1$.

Thermal dust emission is simulated here using the following strategy: we use
`Model 8' of \citet{finkbeiner_etal_1999} in order to extrapolate the
combined COBE-DIRBE and IRAS dust template of \citet{schlegel_etal_1998} to
65~GHz where a comparison with WMAP can be performed.  We then assume a
constant dust polarization fraction $p=11\%$ across our patch.  
This value is in broad agreement with measurements from Archeops which found
$p\simeq5\%$ at low galactic latitudes \citep{benoit_etal_2004,ponthieu_etal_2005},
taking into account the fact that less depolarization along the line of sight is expected at
higher galactic latitudes \citep{prunet_etal_1998}. 
The model adopted here agrees also, as far as the slope and amplitude is concerned,
with the WMAP results based on the large-scale diffuse polarized foreground emission at intermediate 
Galactic latitudes and as summarised in equation~(25) of \citet{page_etal_2007}.
The polarization angle is given by the WMAP dust template 
on large angular scales \citep{page_etal_2007, kogut_etal_2007} 
 and we add in small
 scale power decaying as $C_{\ell}\simeq 45\times\ell^{-3}\mu$K$^2_{\rm CMB}$ for the $E$ and $B$ mode
spectra.  In order to extrapolate the dust from 65GHz back up to our frequency range we
 follow \citet{eriksen_etal_2006} in using `Model 3' of
 \citet{finkbeiner_etal_1999},
\begin{eqnarray}
s_{\textrm{d}}(\nu) = A_{\textrm{d}}
\frac{\nu}{\exp{\frac{h\nu}{kT_{\rm d}}}-1} \frac{\exp{\frac{h\nu_{0}}{kT_{\rm d}}}-1}{\nu_{0}}
\left(\frac{\nu}{\nu_{0}}\right)^{\beta},
\label{eqn:simpScale}
\end{eqnarray}
with $T_{\rm d}=18.1$K and $\beta=1.65$, assumed constant on the patch.  In
Figure~\ref{fig:TQrms} we plot the frequency scaling of the CMB and
dust RMS for total intensity and polarization. The total intensity observations
are characterised by higher signal-to-noise and higher CMB-to-dust ratios than
polarization observations. 

\section{Maximum likelihood problem for component separation -- basic formalism}

\label{sect:mlBasics}

We will model the multi-frequency sky signal as,
\begin{eqnarray}
\bd{d}_p = \bd{A}_p\,\bd{s}_p\,+\,\bd{n}_{p}.
\label{eqn:dataModelIdeal}
\end{eqnarray}
Here,
$\bd{d}_p$ is a data vector containing the measured or estimated
signals for all $n_f$ frequencies and $n_s$ Stokes parameters, which 
are to be analyzed simultaneously;
$\bd{s}_p$ is a vector of the underlying true, and thus to be estimated 
from the data, values of $n_s$ Stokes parameters for each of the $n_c$
components;
$\bd{A}_p \equiv\bd{A}_p\l(\bd{\beta}\r)$ is a component
`mixing matrix', which hereafter will be assumed to be parameterised by
a set of unknown parameters $\l\{\bd{\beta}_i\r\}$.  The mixing
matrix is in general rectangular as the number of components can not
be larger than the number of the frequency channels for which the data
are available. 
In the examples discussed below, we allow only for one unknown spectral 
parameter, the dust slope index, $\beta$, fixing the dust temperature.
Consequently, the elements of the mixing matrix, $\bd{A}$, corresponding
to the dust component, are given by (equation~\ref{eqn:simpScale}), 
\begin{eqnarray}
\bd{A}_{\nu, \rm{dust}} = {s_d\l(\nu, \beta\r)\over s_d\l(\nu_0, \beta\r)},
\end{eqnarray}
for each considered frequency band $\nu$, where $\nu_0$ stands for a  
fiducial, reference frequency set here to $150$GHz. The expression
for the CMB column of the matrix $\bd{A}$ is analogous but does not involve any unknown parameters, 
as its frequency scaling is assumed to be perfectly that of the black-body 
(in antenna units).

With these definitions and models in hand we can write a single pixel
log-likelihood which up to an irrelevant constant  is given by,
\begin{eqnarray}
-2\,\ln\,{\cal L}\l(\bd{s}_p, \bd{\beta}_i\r) = \const +
\l(\bd{d}_p-\bd{A}_p\,\bd{s}_p\r)^t\,\bd{N}_p^{-1}\l(\bd{d}_p-\bd{A}_p\,\bd{s}_p\r).
\label{eqn:genSpixLike}
\end{eqnarray}
Here, $\bd{N}_p$ is a square, symmetric noise matrix of the frequency
maps for a pixel $p$, with rank $n_s\times n_f$. This likelihood is
the basis of the \citet{eriksen_etal_2006} approach.  It is clearly
straightforward to introduce a multi-pixel version of this
likelihood, and with the definition as above we have,
\begin{eqnarray}
-2\,\ln\,{\cal L}_{data}\l(\bd{s}, \bd{\beta}\r) = \const +
\l(\bd{d}-\bd{A}\,\bd{s}\r)^t\,\bd{N}^{-1}\l(\bd{d}-\bd{A}\,\bd{s}\r),
\label{eqn:genMpixLike}
\end{eqnarray}
where all the matrices and vectors now span over many
pixels. Hereafter, we will refer to this likelihood as the full data
likelihood.  In the simple case of the matrix  $\bd{N}$ being
block-diagonal, thus allowing only for correlations between different Stokes
parameters at each pixel, equation~(\ref{eqn:genMpixLike}) simplifies to read,
\begin{eqnarray}
-2\,\ln\,{\cal L}_{data}\l(\bd{s}, \bd{\beta}\r) = \const +
\sum_p\,\l(\bd{d}_p-\bd{A}_p\,\bd{s}_p\r)^t\,\bd{N}_p^{-1}\l(\bd{d}_p-\bd{A}_p\,\bd{s}_p\r).
\label{eqn:genMpixLikeNoCorr}
\end{eqnarray} 
This likelihood reaches its maximum for the values of $\bd{s}$ and
$\bd{\beta}$ fulfilling the relations,
 \begin{eqnarray}
-\l(\bd{A}_{,\bd{\beta}}\,\bd{s}\r)^t\,\bd{N}^{-1}\,\l(\bd{d}-\bd{A}\,\bd{s}\r) = 0
\label{eqn:maxRels}
\\
\bd{s} = \l(\bd{A}^t\,\bd{N}^{-1}\,\bd{A}\r)^{-1}\,\bd{A}^t\,\bd{N}^{-1}\,\bd{d},
\label{eqn:step2amps}
 \end{eqnarray}
where $_{,\bd{\beta}}$ denotes a partial derivative with respect to
$\bd{\beta}_i$.  Solving this system of equations above can be problematic, 
in particular in the case of multiple spectral parameters, $\l\{\bd{\beta}_i\r\}$, 
due to their generally non-linear character \citep{brandt_etal_1994,
eriksen_etal_2006}.  However, once the values of the spectral
parameters are found the second set of linear equations provides
straightforward estimates of the pixel amplitudes.
Equations~(\ref{eqn:genMpixLike}), ~(\ref{eqn:genMpixLikeNoCorr}), \&\
(\ref{eqn:step2amps}), are the starting point for our discussion in the following.

\subsection{Constraining the spectral parameters}

\label{sect:specParams}

\subsubsection{Maximum likelihood estimator}

\label{sect:speclike}

Equation~(\ref{eqn:step2amps}) provides an estimate of the sky components
for any assumed set of spectral parameter values, $\bd{\beta}$. This estimate
maximizes locally the full data likelihood restricted to
a subspace of the parameter space as defined by the assumed values of the spectral parameters.
We can use that equation to eliminate the sky signals from the full data likelihood 
expression, equation~(\ref{eqn:genMpixLike}), obtaining,
\begin{eqnarray}
-2\,\ln {\cal L}_{spec}\l(\bd{\beta}\r) & = & \hbox{{\sc const}} - \l( \bd{A}^t\,\bd{N}^{-1}\,\bd{d}\r)^t\,\l(\bd{A}^t\,\bd{N}^{-1}\,\bd{A}\r)^{-1} \l( \bd{A}^t\,\bd{N}^{-1}\,\bd{d}\r).
\label{eqn:slopelikeMpixUnbiased}
\end{eqnarray}
The above expression permits a calculation of the `ridge' values of the full data likelihood, i.e., the maximum
value of the likelihood as attained for any chosen value of the spectral parameters and, therefore, is maximized
by the true (global) maximum likelihood value of the spectral parameters. Hereafter, we will call this function a
spectral likelihood and use it to determine the maximum
likelihood estimates of the spectral parameters and obtain an insight into the resulting errors.
We point out that the spectral likelihood is not a proper likelihood in a common sense, and is thus often
referred to as a `ridge' or `profile' likelihood to emphasize this fact \citep[e.g.,][]{nielsen_cox_1994}. Nevertheless,
it permits to retrieve the maximum likelihood values of the spectral parameters,
and provides useful, if not strictly precise, estimates of their uncertainty.

In the alternate approach as implemented by \citet{eriksen_etal_2006} the spectral parameters are estimated with 
help
of the marginal distribution derived for the full data likelihood, equation~(\ref{eqn:genMpixLike}). The marginalisation 
over the sky amplitudes is performed numerically, via MCMC, and assumes flat priors for the sky signals.  
We will show here that that approach is not only more computationally involved but also fails to reproduce
 the maximum 
likelihood estimates of the spectral parameters derived by maximizing equation~(\ref{eqn:genMpixLike}).

\begin{figure} 
\begin{center}
\includegraphics[height=2in, angle = 0]{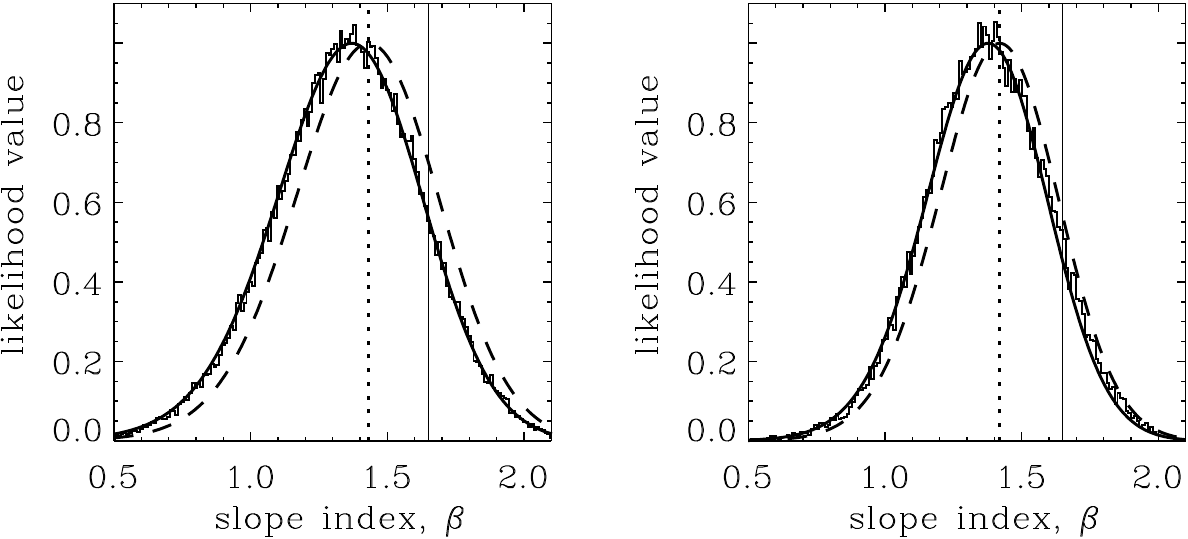}
\caption{ The one dimensional likelihoods for the slope index,
$\beta$, with the sky component amplitudes marginalised over and
computed for the total intensity data and the Stokes $Q$ parameter
({\bf left and right panels}). The histograms
 show the result of the MCMC sampling of the full data
likelihood, equation~(\ref{eqn:genMpixLike}). The thick solid and 
dashed lines show the result for the marginal and spectral
 likelihood expressions,
equations~(\ref{eqn:slopelikeMpixGen}) and (\ref{eqn:slopelikeMpixUnbiased}) 
respectively. The dotted vertical lines show the $\beta$ value
corresponding to a peak of the full data likelihood
equation~(\ref{eqn:genMpixLike}), found through a direct grid search.
The displayed results are calculated for a single pixel with $n_{side} = 1024$
for the total intensity, and $n_{side} = 16$ for the polarization. 
The true value of the slope index, $\beta = 1.65$, is indicated by the solid vertical lines.
\label{fig:slopeMarg1dimNoOff}
}
\end{center}
\end{figure}

We first observe that the marginalisation procedure can be done analytically and that the integration over the
sky amplitudes leads to,
\begin{eqnarray}
-2\,\ln {\cal L}_{marg}\l(\bd{\beta}\r) & = & 
- 2\,\ln \int\,d\,\bd{s}\,\exp\l[ -{1\over 2}\, \l(\bd{d}-\bd{A}\,\bd{s}\r)^t\,\bd{N}^{-1}\l(\bd{d}-\bd{A}\,\bd{s}\r)\r]
 \nonumber \\
& = & \hbox{{\sc const}} - \l( \bd{A}^t\,\bd{N}^{-1}\,\bd{d}\r)^t\,\l(\bd{A}^t\,\bd{N}^{-1}\,\bd{A}\r)^{-1}
\l( \bd{A}^t\,\bd{N}^{-1}\,\bd{d}\r) + \ln \l| \l(\bd{A}^t\,\bd{N}^{-1}\,\bd{A}\r)^{-1} \r|,
\label{eqn:slopelikeMpixGen}
\end{eqnarray}
where  $\l|...\r|$
denotes the matrix determinant. This expression defines the likelihood
function for the parameters, $\bd{\beta}$, given the data, $\bd{d}$,
assuming an explicit independence of any spectral parameter, $\bd{\beta}$,
on $\bd{s}$. 
(We remark here in passing that though this equation is derived under the assumption of the flat 
priors for the component amplitudes. However, any kind of a spatial template-like prior constraint with
known, and Gaussian, uncertainty can be easily incorporated here, and treated as yet another
data set in equation~(\ref{eqn:genMpixLike}).) In  Figure~\ref{fig:slopeMarg1dimNoOff} we show that
indeed our analytical result for the marginal likelihood reproduces perfectly an outcome of the numerical marginalisation (via MCMC) of the full data likelihood.

The marginalised and spectral likelihoods, 
equations~(\ref{eqn:slopelikeMpixGen}) \&~(\ref{eqn:slopelikeMpixUnbiased}), clearly differ.
As the difference depends on the spectral parameters, we can conclude that the marginal likelihood peaks at their
values, which do not coincide with their maximum likelihood estimates. We can see this more 
explicitly by calculating the first derivative of equation~(\ref{eqn:slopelikeMpixGen}),
\begin{eqnarray}
-2 {\partial \,\ln {\cal L}_{marg}\over \partial \bd{\beta}_i} = & - & 2\,\l(\bd{A}_{,\bd{\beta_i}}^t\,\bd{N}^{-1}\,\bd{d}\r)^t\,\bd{\hat{N}}\,\l(\bd{A}^t\,\bd{N}^{-1}\,\bd{d}\r)
+ 2\,\l(\bd{A}^t\,\bd{N}^{-1}\,\bd{d}\r)^t\,\bd{\hat{N}}\,
\bd{A}^t\,\bd{N}^{-1}\,\bd{A}_{,\bd{\beta}_i}
\,\bd{\hat{N}}\,\l(\bd{A}^t\,\bd{N}^{-1}\,\bd{d}\r)\nonumber\\
& - & 2 \, {\rm tr} \, \l[ \bd{\hat{N}}\,\l(\bd{A}^t\,\bd{N}^{-1}\,\bd{A}_{,\bd{\beta}_i}\r)\r],
\label{eqn:slopelikeMpixGenDer}
\end{eqnarray}
which, in general, does not vanish for the maximum likelihood values of the spectral indices. Indeed, we have for $\bd{\beta} = \bd{\beta}_{ML}$, as defined by equations~(\ref{eqn:maxRels}) \&~(\ref{eqn:step2amps}),
\begin{eqnarray}
-2 \, \l. \,{\partial \, \ln {\cal L}_{marg}\over \partial \bd{\beta}_i}\r|_{\bd{\beta} = \bd{\beta}_{ML}}
= - 2 \, \l. {\rm tr} \, \bd{\hat{N}}\,\bd{A}^t\,\bd{N}^{-1}\,\bd{A}_{,\bd{\beta}_i}\r|_{\bd{\beta} = \bd{\beta}_{ML}}.
\end{eqnarray}

The two major observations that we have made in this Section are visualized in 
Figure~\ref{fig:slopeMarg1dimNoOff}.
It shows the presence of a mismatch between the maximum 
likelihood value of the slope index derived from the marginalised likelihood and that obtained from
the full data likelihood, as well as a lack of it if the spectral likelihood, equation~(\ref{eqn:slopelikeMpixUnbiased}), is utilized, instead of the marginalised one. Note that all these effects, though small in the presented examples 
due to our assumptions, are however clearly discernible. 
We present a more detailed discussion of the numerical results in the next Section.

Deriving the best possible constraints on a set of the
spectral parameters requires using in the likelihood analysis all
the available data points for which the frequency scaling of the component
is, or can be assumed the same. When the scaling is identical for all
the data analyzed, the mixing matrix $\bd{A}$ is made of identical,
diagonal blocks, a fact which can be additionally exploited to further
speed up the numerical calculations and/or simplify the implementation
of the formalism.  However, the expressions derived above are fully
general and can be applied in the cases with arbitrary mixing
matrices 
for instance, such as those corresponding to multiple sets
of the spectral parameters, each describing the frequency scaling of
different subsets of all pixels. All the scaling parameters are then
estimated simultaneously, and we will use this fact later in the
paper.
 
 \begin{figure}
\begin{center}
\includegraphics[height=2.0in, angle = 0]{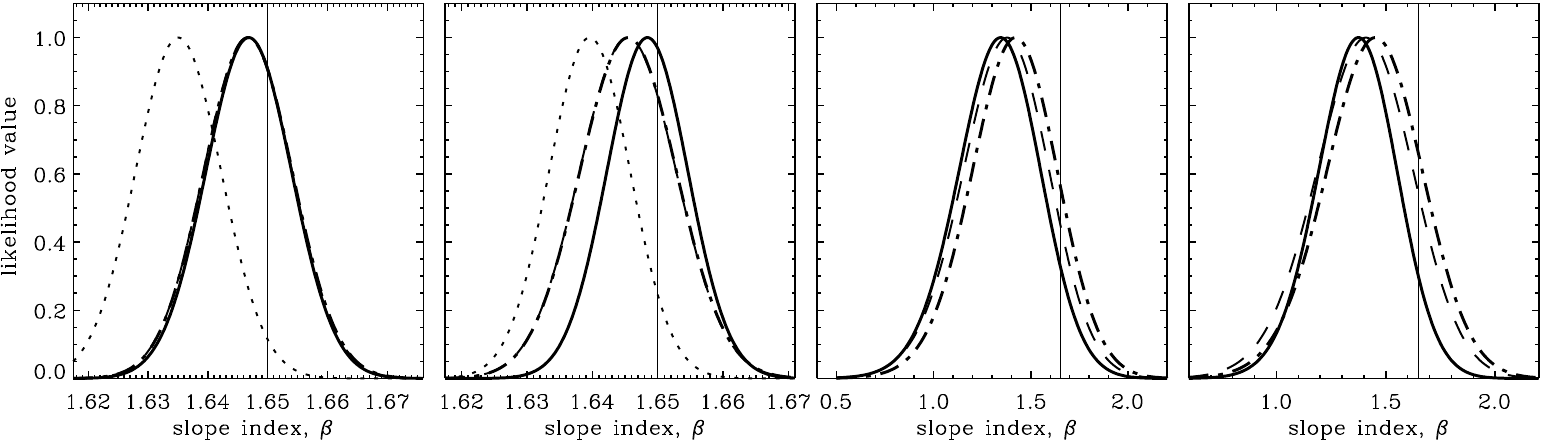}
\caption{ The one dimensional likelihoods for the slope index,
$\beta$, with the sky component amplitudes marginalised over and
computed for the total intensity  and  Stokes $Q$ parameter data
({\bf two left and two right panels}). 
The results shown here are from the analysis involving $4096$ high resolution pixels.
The solid and dotted lines (the latter missing in the $Q$ data case, see text) 
represent the spectral and marginalised likelihood cases,
equations~(\ref{eqn:slopelikeMpixUnbiased}) and (\ref{eqn:slopelikeMpixGen}) respectively, 
computed using the {\em multi-pixel approach}. The 
dashed and dot-dashed lines (nearly perfectly overlapping)
show the corresponding results derived in the {\em single-pixel approach}. 
The homogeneous and inhomogeneous noise cases are shown by the first and second panel,
respectively, of each pair of panels.
The true value, $\beta= 1.65$, is indicated by the solid vertical lines.
\label{fig:slopeMarg1dimNoOff_discussion}}
\end{center}
\end{figure}
 
 \subsubsection{Comparison and examples}
We can apply the formulae presented here without any preprocessing to
 subsets of the available maps for which the assumption of a single set of
 spectral parameters is physically justifiable. Alternately, as proposed by
 \citet{eriksen_etal_2006} we could first smooth the available data and
 downgrade it to a lower resolution, prior to calculating the spectral
 likelihood for each low-resolution pixel separately. The spectral
 parameters determined in this way are then applied to all the high
 resolution pixels falling into the low resolution one.  Hereafter, we refer
 to these two implementations of the formalism as a {\em multi-pixel} and
 {\em single-pixel} approach, respectively.

 Figure~\ref{fig:slopeMarg1dimNoOff_discussion} shows the one
 dimensional likelihood for the spectral index $\beta$ -- both the marginalised, equation~(\ref{eqn:genMpixLike}), 
 and spectral, equation~(\ref{eqn:slopelikeMpixUnbiased}), renditions as computed in the
 single- and multi-pixel approaches.
 The two left panels show the total intensity cases for homogeneous and inhomogeneous noise
 distribution.
 The latter was obtained by allowing for a $\sqrt{3}$ noise RMS variation
across the patch, characterised by the same value on zones with size
of about $14'$ ($n_{\rm side}=256$), but changing randomly from
zone to zone.
The solid and dotted lines represent the
spectral and the marginalised likelihood, respectively in the multi-pixel
approach, while the nearly overlapping dashed and dot-dashed lines are
for the single-pixel analysis. 
In the inhomogeneous noise
case, the difference between the single- and multi- pixel approaches
as well as marginal and spectral likelihoods is now much more evident
and manifests itself as an overall shift and broadening, showing the
macroscopic gain in adopting the expression in
equation~(\ref{eqn:slopelikeMpixUnbiased}) throughout the analysis.

The two right panels of Figure~\ref{fig:slopeMarg1dimNoOff_discussion} 
show analogous cases but for the
Stokes $Q$ parameter. The marginalised multi-pixel
likelihoods (dotted lines) are, however, not shown as they do not fit
the $x$-axis ranges.  Note that unlike in the total intensity case the
single and multi-pixel likelihoods differ even in the homogeneous
noise case (middle panel, the solid and dashed lines). This
reflects the fact that in the simulations used here there is more small-scale power in the dust
polarization than in the total intensity, which leads to the loss of
constraining power as a result of smoothing in the single pixel
approach (see the discussion below). The effect is
additionally enhanced if the inhomogeneous noise is present as shown
in the right panel. In both cases the mismatch and the loss of precision are small, as
expected given a relatively smooth variation of the dust foreground as
adopted here.

\begin{figure}
\begin{center}
\includegraphics[height=2.0in, angle = 0]{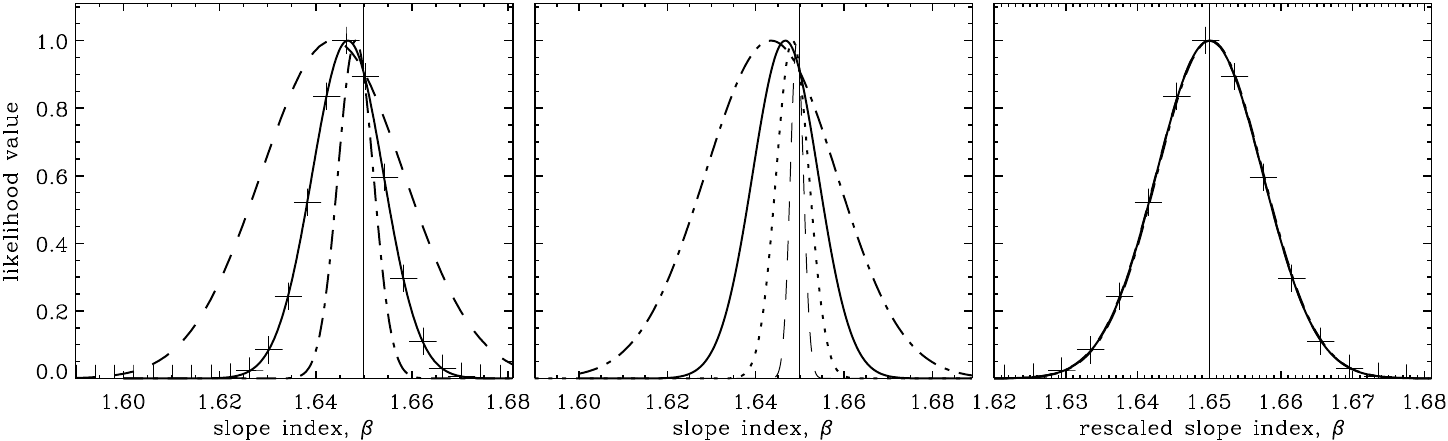}
\caption{The likelihood constraint on the slope index for different
noise levels and dust amplitudes.  The {\bf left panel} shows the spectral
likelihoods calculated for a single pixel and three different noise
levels: $50\%,\ 100\%$ and $200\%$ of the nominal noise value
corresponding to the low resolution total intensity pixel as described
in Section~\ref{section:dataSet}.  They are depicted with dot-dashed,
solid and dashed lines respectively. In all three cases the
same sky signal amplitudes for all components are assumed. In
addition, the crosses correspond to the case with both the noise
levels and the dust amplitude amplified by a factor $2$.  The {\bf middle}
panel shows the effects of changing the level of the dust signal level in the
input data, while keeping all the other parameters constant. The displayed lines
correspond to the cases with the dust amplitude rescaled by a factor of: $0.5$ (dot-dashed),
$1$ (solid), $2$ (dotted), and $5$ (dashed).
 The {\bf right
panel} shows all the cases from the two previous panels but now rescaled by a factor proportional to
a ratio of the {\rm dust\ amplitude} and the {\rm RMS noise\ level} 
 and shifted to have a peak at the true value of $\beta
=1.65$, indicated in all panels with a vertical line. \label{fig:errorScaling}}
\end{center}
\end{figure}

The uncertainty involved in the determination of the spectral parameters
depends on the noise level and the amplitude of the sky components. It is
visualised for a single pixel in Figure~\ref{fig:errorScaling}, where the
noise and dust amplitudes are rescaled in order to check their impact on the
precision of the determination of $\beta$. In the left panel, the noise RMS
and the actual noise level are varied as explained in the caption, while in the 
middle panel, only the dust amplitude is rescaled. In both cases the 
width of the likelihood changes accordingly.
The right panel shows the 
same likelihoods shown in the two other panels but rescaled by a factor 
proportional to the ratio of the RMS of 
the dust and noise, and shifted to have the maximum at $\beta=1.65$.  
In these examples the constraints on the slope parameter are found independent
of the level of the CMB anisotropy. This is because the frequency
scaling of the CMB component is assumed to be known a priori.
The increase of the actual noise level in a given pixel, with the noise RMS
kept constant, leads to an overall shift of the likelihood peak, thus changing 
(and thus potentially biasing) the best-fitting $\beta$ value, but does not affect 
the likelihood width.

By using more
pixels, the constraints can be usually improved. The rate at which that
takes place, will however depend on the specific, and not known a
priori, magnitude of the sky component in the newly added pixels, as well
as, pixel instrumental noise levels, and therefore may not conform
with the usual, $\sim 1/\sqrt{n_{pix}}$, expectation. Nevertheless, for the
diffuse components, and in particular at high Galactic latitudes, as for
example the dust foreground considered in the examples presented here, we
find that once the number of included sky pixels is large enough the
uncertainty of the slope determination falls roughly as expected, i.e.,
$\sim 1/\sqrt{n_{pix}}$. However, even in an extreme case when the newly
added pixels happen to contain no information about a given component, the
uncertainty, estimated using the multi-pixel approach, is guaranteed not to
deteriorate.  This may not be however the case with the single pixel
approach, where including too many, for instance, dust-free pixels may
suppress the noise power less than that of the dust and consequently
increase the errors of the dust parameter determination.  In a less extreme
and more common case, the smoothing of the rapidly varying sky components
whose amplitudes change across the low resolution pixel, will somewhat
affect the precision of the spectral parameter determination, though will
not bias the estimation result. The presence of the inhomogeneous noise on
the scales smaller than the low resolution pixel will also have similar
consequences (Figure~\ref{fig:slopeMarg1dimNoOff_discussion}). 

Nevertheless, in the examples considered here, and concurred by those
of \citet{eriksen_etal_2006}, the smoothing with a progressively larger window 
generally leads to an improvement of the constraints on the spectral parameters.  
This is because as the result of the smoothing the noise is typically
suppressed more significantly than the dust component. In all the
cases, the multi-pixel approach will produce nearly optimal
constraints, independently of the actual sky distribution of the
signals.  It is also more
flexible as it permits arbitrary pixel subsets for which identical
spectral parameter values are assumed, and thus can better deal with
the masked pixels and/or sky patch edge effects. Moreover, as discussed
before, it also treats more optimally cases in which either the noise or the 
relative component content varies rapidly across the sky.
For all these reasons the multi-pixel approach is therefore an approach 
of the choice here.

The evaluation of the spectral index is the first step toward the
recovery of the actual maps of the components superposed in the data,
with the latter step treated in the next Section.  Before concluding
this part, however, we draw a few more comments on the numerical
aspects of the likelihood treatment presented here, as well as on the
relation between pixel size and constraint on the spectral index.

\subsubsection{Numerical issues}

The spectral likelihood
equation~(\ref{eqn:slopelikeMpixUnbiased}) is more amenable to a numerical
work than the marginalised expression in equation~(\ref{eqn:slopelikeMpixGen}), since
it does not require computation of a matrix determinant.  In fact, it
can be evaluated in a numerically efficient way, requiring at most on
order of ${\cal O}(n_s^2\,n_f^2\,n_{pix}^2)$ floating point operations
and often many fewer.  For instance, the potentially time consuming,
inversion of the matrix, $\bd{A}^t\,\bd{N}^{-1}\,\bd{A}$, involved in
its computation can be sometimes simplified by exploiting its
particular structure, for instance, block-diagonality, if the data
noise $\bd{N}$ is diagonal.  Alternately and more generally, the
computation of the inverse can be replaced by solving a linear set of
equations and thus performed using fast, iterative techniques, e.g.,
preconditioned Conjugate Gradient \citep{Golub_book, shewchuk1994},
which involve only matrix-vector products. These can be efficiently
done even if the noise in the sky map is correlated, a usual feature
of the small-scale experiments.  We emphasise that for many
forthcoming CMB experiments even an explicit calculation of the full
noise covariance matrix for the input maps $\bd{N}$ may be already
unfeasible.  The computations involved in the evaluation of the
spectral likelihood should be ultimately related to the time domain
data as gathered by the CMB experiments.  This can be done by
exploiting techniques similar to those employed on the map-making
stage of the CMB data analysis \citep[e.g.,][]{ashdown_etal_2007} and
combining them with the formulae presented here. We leave the further
development of the algorithm described here along these lines for
future work, restricting hereafter ourselves to the cases when the
noise correlation matrix $\bd{N}$ is diagonal, and the calculations
are done in the pixel domain. In such a case the numerical workload is
brought down to merely ${\cal O}\l(n_{pix}\r)$.

\subsubsection{Pixel size and constraints on the spectral parameters}

The r\^ole of a pixel and, in particular, its size may appear
more difficult to grasp in the context discussed here than in, for
example, the standard map making procedure. The pixel size determines
the number of input data points and at the same time the number of
independent degrees of freedom which are subsequently marginalised
over. It may however, or may not, affect the number of spectral
parameters which are to be constrained on this step.

Using too crude a pixelization may compromise the quality of the
determination of the spectral parameters whenever the instrumental
noise is inhomogeneous and the sky signal changes on the pixel scale
(Figure~\ref{fig:slopeMarg1dimNoOff_discussion}).  The
pixel size should be therefore selected small enough to make such a
variation negligible.  This is in fact a usual criterion used in the
production of the CMB maps, where the pixel size is usually fixed as
some fraction of the resolution of the instrument.  Also as in the
map-making case, there is no loss of precision due
overpixelization, if only we constrain the mixing matrix to be the
same for all sub-pixels and thus we do not introduce any extra scaling
parameters. That can be seen directly from
equation~(\ref{eqn:slopelikeMpixUnbiased}) by noting that for all the
sub-pixels of a given pixel, the sky signal is in fact the same, even
though we introduce multiple parameters, i.e., sub-pixel amplitudes,
to describe them.  Clearly, in such circumstances overpixelizing does
not lead to any gain and may affect a performance of the numerical
calculations by unnecessarily exaggerating the size of the problem.

To conclude this discussion we note that for polarized map-making
the pixel size may need to be fixed to alleviate the total intensity
leakage to the $Q$ and $U$ Stokes parameter maps. This may result in a
pixel size which is unnecessarily tiny for the purposes discussed
here. In such a case, it may be useful to downgrade the pixelization
during the component separation procedure, an operation which, in the
formalism presented here, can be done by introducing a single set of
sky component amplitudes for multiple pixels of the input maps.

\subsection{Constraining the component amplitudes}

\label{sect:compAmpDet}

We can calculate the maximum likelihood estimates of the component
maps, given the maximum likelihood estimates of the spectral
parameters derived in the previous section and expression for the peak
of the data likelihood, equation~(\ref{eqn:step2amps}). The two step
algorithm, which incorporates these two procedures, is therefore a
genuinely maximum likelihood approach, as we set to develop in this
paper. We note that equation~(\ref{eqn:step2amps}) is commonly employed in
the component separation techniques 
and in particular, it also defines the second step in the approach of
\citet{eriksen_etal_2006}.  However, given that the spectral estimates
which those methods obtained on the first step do not coincide with,
or can not be straightforwardly related to, the maximum likelihood
values, the recovered sky signal values correspond only to a
likelihood ridge value (a maximum of the likelihood for given values
of the spectral parameters) rather than a global maximum.

\begin{figure}
\centerline
{
\includegraphics[height=1.75in, angle = 0]{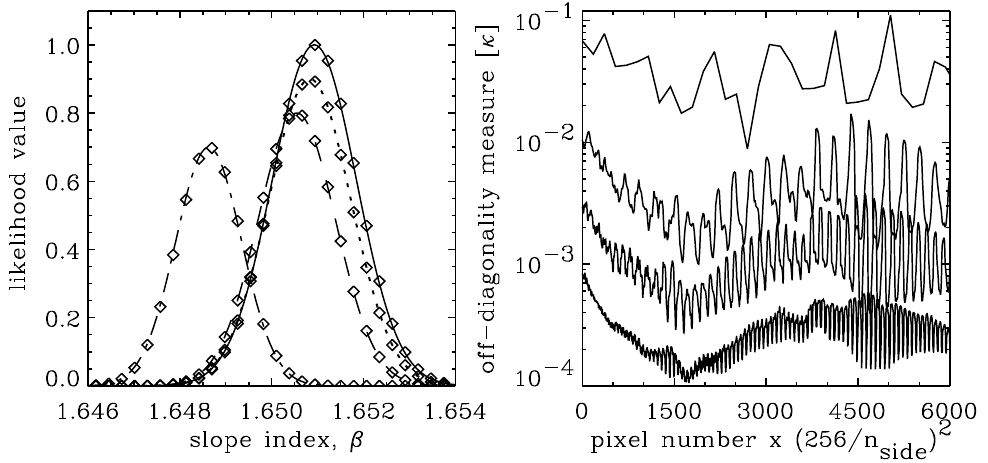}
\includegraphics[height=1.75in, angle = 0]{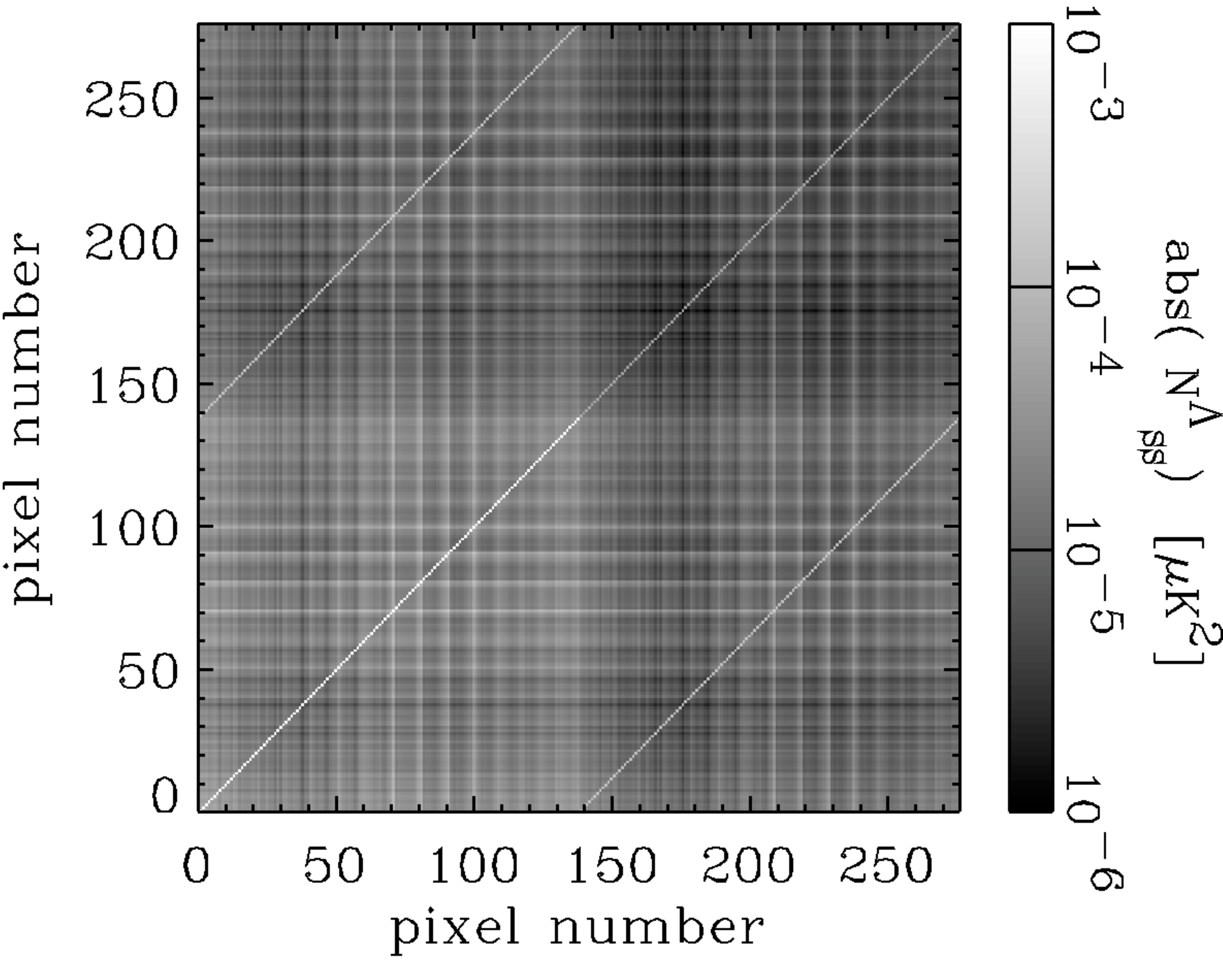}
}
\caption{{\bf Left panel}: The likelihoods for the slope index, $\beta$, computed using equation~(\ref{eqn:slopelikeMpixUnbiased}) 
for the total intensity maps and 4 different resolutions: $n_{side} = 256, 128, 64, 16$, depicted with solid, dotted, dashed
and short-long dashed lines respectively. The squares show the Gaussian approximation to each of the likelihoods
with RMS computed using the curvature matrix formalism, equation~(\ref{eqn:sRMS}). The likelihoods were rescaled for the visualization
reasons.
{\bf Middle panel:} The non-diagonality
measure, $\bd{\kappa}_{pp}$, as a function of a pixel number for the same four data renditions as in the left panel. The resolution
decreases from bottom-up and the three bottom curves have been somewhat smoothed to facilitate a better comparison. {\bf Right panel:} The covariance matrix for CMB and dust components recovered from the
total intensity data with the pixel size $n_{side} = 64$. The two diagonal blocks correspond to CMB--CMB (bottom left), 
dust--dust (top right), while two off-diagonal blocks show their cross-correlation. 
\label{fig:curvMat}}
\end{figure}

Equation~(\ref{eqn:step2amps}) also corresponds to a standard general least
square solution for the components, if the spectral parameters are
perfectly known ahead of the time. 
In this case the solution error is
quantified by the error correlation matrix $\bd{\hat{N}}$, given by,
\begin{eqnarray}
\bd{\hat{N}} \equiv \l(\bd{A}^t\,\bd{N}^{-1}\,\bd{A}\r)^{-1}.
\label{eqn:noiseCorrOptim}
\end{eqnarray} 
Whenever the spectral parameters are
not known perfectly, and need to be constrained from the data, we can
get some insight into the structure of the error correlation matrix
from the curvature matrix computed at the peak of the likelihood. The
relevant formulae are listed in  Appendix~\ref{app:curvFullLike}. 
The results of their application to our test cases  are visualised in Figure~\ref{fig:curvMat}.

We first note that the Gaussian approximation to the spectral likelihood of the slope
index with the RMS error derived from the curvature matrix formalism (equation~\ref{eqn:sRMS})
fares extremely
well in reproducing the actual likelihood as calculated using equation~(\ref{eqn:slopelikeMpixUnbiased}). 
We show
the comparison in the left panel of Figure~\ref{fig:curvMat}. The four likelihoods shown
there correspond to four different pixel sizes used to represent the same input data. 
The respective RMS uncertainties and peak positions are therefore very similar with
small differences attributable to the presence of the small scale power as discussed
earlier.
In all cases the agreement between the approximate (squares) and exact computation
is very good. This supports the expectation that the curvature matrix approach can be
also successful in describing the `linear' degrees of freedom such as the component maps.

The respective covariance matrix, $\bd{N^\Lambda}_{\bd{s}\,\bd{s}}$, (equation~\ref{eqn:noiseCorrFromCurvMat})
for the recovered component maps  is shown in the right panel
of Figure~\ref{fig:curvMat}. The four blocks shown in there correspond to the CMB--CMB (bottom left),
dust--dust (top right) and CMB--dust and dust--CMB (bottom right and top left respectively)
cases. The two most salient features of the matrix are the strong correlations between the amplitudes
of different recovered components in any single sky pixel, as reflected by the prominent diagonals of the
off-diagonal blocks displayed in the figure, and the presence of the off-diagonal correlations
for each of the two components. The former feature is already encountered in the cases in which
all the spectral parameters are known precisely and whenever the matrix $\bd{A}^t\bd{A}$
happens to be non-diagonal. It simply reflects the fact that the estimates are derived from the same
input data. However,  if the spectral parameters are also recovered from the data, and therefore, are
 known only with a limited precision, these may lead to a leakage between different sky component signals,
 resulting in the presence of the off-diagonal correlations between the errors with which the 
 component amplitudes can be estimated. Both these are quantified by the expression
 for the covariance matrix of the sky component amplitudes in equation~(\ref{eqn:noiseCorrFromCurvMat}). 

 The tartan-like pattern seen in  the right panel of Figure~\ref{fig:curvMat} reflects the fact that the correction 
 term in this equation has a form of a product of two low-rank matrices, with the brightest, vertical and
 horizontal stripes correspond to the pixels with the largest dust signal amplitude. 
This highlights the fact that the second `correction' term on the rhs of equation~(\ref{eqn:noiseCorrFromCurvMat}) 
depends on two factors:
the error on the spectral parameters, and the amplitude of the actual sky signals. In the test cases 
 considered here, we find the error correlation matrix, 
 $\bd{N^\Lambda}_{\bd{s}\,\bd{s}}$, is strongly diagonal-dominated for the renditions of the data with small
 pixels but is becoming progressively less diagonal when the pixel-size increases. This is illustrated
 in the middle panel of Figure~\ref{fig:curvMat}, where we show a non-diagonality measure, $\bd{\kappa}$, 
 (defined in Appendix~\ref{app:matOff}) for all considered pixels and different pixel resolutions,
  to quantify the importance of the off-diagonal elements of the matrix, $\bd{N^\Lambda}_{\bd{s}\, \bd{s}}$ and, 
  therefore of the correction term in equation~(\ref{eqn:noiseCorrFromCurvMat}). Whenever $\bd{\kappa} \ll 1$, the off-diagonal elements
 are likely to be unimportant, and then the matrix $\bd{\hat{N}}$ may provide a good approximation for the 
 overall pixel-pixel correlation pattern. If, however, $\bd{\kappa} \simlt 1$, the off-diagonal elements, and thus
 the correction term in equation~(\ref{eqn:noiseCorrFromCurvMat}) may need to be taken into account.
 We see that for our multi-resolution rendition of the data the transition between the two regimes happens
 to be somewhere around $n_{side} \sim 64$. This suggests that though simple uncorrelated noise  model (for 
 each component separately), as implied in this case by the matrix $\bd{\hat{N}}$, may be a sufficient 
 description of the component map error for high-$\ell$ part of the power spectrum, however, at its lower end 
 -- a more sophisticated treatment may be needed.
 
 We point out that the increase in the importance of the non-diagonal correction term with the growing
 size of the pixels is not induced by the change in the error in the determination of the spectral index,
 as this remains roughly constant (see the left panel of Figure~\ref{fig:curvMat}), but is due to the increasing signal-to-noise
 ratio of the component signals. We stress that the case discussed here is rather extreme,
 as it treats the total intensity signal with the noise level as expected for the experiments targeting the
 $B$ mode polarization. In the cases with the lower signal-to-noise across the range of scales, the off-diagonal
 elements may well be less important even at the largest angular scales considered. Indeed, for the
 $Q$ Stokes parameter maps considered in this paper the diagonal approximation seems to be breaking down 
 only at the lowest resolution ($n_{side} = 16$). The significance of the off-diagonal elements will be,
 however, enhanced whenever more spectral parameters is considered, as this will unavoidably boost 
 the spectral parameter determination uncertainty. This could be,  for instance, the case when a multiple
 sets of the spectral parameters are introduced to account for their spatial variability, or whenever the
 calibration uncertainty  of the input maps has to be included in  the analysis. 
 We discuss this latter case in more detail later.
 
 We also emphasise that the question whether the covariance estimate
 based on the curvature matrix,
 equation~(\ref{eqn:noiseCorrFromCurvMat}), provides a sufficiently
 good approximation to the actual covariance of the recovered sky
 components may in fact need to be re-evaluated
 application-by-application.
 
From the numerical point of view, as in the context of the
marginalised likelihood for the spectral parameters, ${\cal L}_{spec}$, we can recast the algebraic operations
involved in the calculations of the component amplitudes, equation~(\ref{eqn:step2amps})
as matrix vector multiplications and which can be performed with help of the iterative solvers. 
The same is true of the correction terms in the expressions for the covariance matrices. The calculation
of the matrix, $\bd{\hat N}$, may however present a significant computational challenge.
As our focus here is on the small-scale experiments 
this problem can be overcome with aid of the available supercomputers.

Summarizing, in this Section we have presented a  likelihood
based estimator of the spectral parameters, equation~(\ref{eqn:slopelikeMpixUnbiased}),
 and shown that it coincides with
the maximum likelihood predictions derived from the maximization of
the full likelihood. The new estimator does not require any
pre-processing, handles optimally the inhomogenous noise cases and
avoids problems with the observed sky edge and masked pixel
effects. Combined with the standard determination of the component
amplitudes, equation~(\ref{eqn:step2amps}), it provides a two-step, maximum
likelihood estimator of the sky components and the spectral
parameters, which is computationally tractable. We also have introduced relevant 
curvature matrices to characterise statistical properties of
the derived component maps and spectral parameters.

\section{Maximum likelihood problem for component separation -- extensions}
\label{sect:mlExts}

In this Section we include extra parameters in the likelihood expression, 
which account for common instrumental characteristics, namely the 
presence of the undetermined offsets in the input data, and calibration 
errors. 

\subsection{Unconstrained single frequency map offsets}

The single frequency maps produced from the data of the most CMB 
experiments have often arbitrary global offsets \citep[e.g.,][]{stompor_etal_2002}, which
for instance, could result from either instrumental $1/f$--noise, differential 
scanning strategy or nearly sky synchronous, systematic effects \citep[e.g.][]{johnson_etal_2007}.

The data model for each map signal may need therefore to be more complex than 
it is assumed in equation~(\ref{eqn:dataModelIdeal}),
\begin{eqnarray}
\bd{d}_{p} = \bd{A}_{p}\,\bd{s}_{p} + \bd{o} + \bd{n}_{p},
\label{eqn:dataModelWoff}
\end{eqnarray}
where, $\bf{o}$ is a vector of $n_f\times n_s$ components, each of which describes an 
offset of a map at a given frequency and for each of the Stokes parameters.
The likelihood function needs to be generalised accordingly,
\begin{eqnarray}
-2\,\ln {\cal L}_{data} \l( \bd{s}, \bd{\beta}, \bd{o}\r) = \hbox{\sc const} + \l(\bd{d}-\bd{A}\,\bd{s}-\bd{o}\r)^t\,\bd{N}^{-1}\l(\bd{d}-\bd{A}\bd{s}-\bd{o}\r). 
\label{eqn:genMpixLikeOff}
\end{eqnarray}
Clearly, as we have just introduced $n_f\times n_s$ extra unknowns,
the likelihood for a single pixel can not be uniquely
maximised. However, as by definition the offset vector, $\bd{o}$, does
not depend on a pixel, one could hope that the problem is becoming
well-posed if a sufficient number of pixels is analyzed
simultaneously. We will show that although such an expectation is not
fully borne out, and a more subtle treatment of the arbitrary map
offsets is needed, meaningful constraints can be derived for the
spectra parameters and the component maps in such a case.

As mentioned above the offsets are completely spurious and consequently do not
contain any cosmologically relevant information, we can therefore marginalise over them
without any loss of information. Assuming the Gaussian priors for all
the offsets, with mean zero and a dispersion given by $\xi_i$, we
arrive at
\begin{eqnarray}
-2\,\ln\,{\cal L}'_{data}\l(\bd{s}, \bd{\beta}\r) & \equiv & -2\,\ln\,\exp \int\,d\bd{o}\,{\cal L}  \nonumber\\
& = & \hbox{\sc const} 
+ \l(\bd{d}-\bd{A}\,\bd{s}\r)^t\,
\l[\bd{N}^{-1} - \l(\bd{N}^{-1}\,\bd{U}\r)\,\l(\bd{\Xi}^{-1} + \bd{U}^t\,\bd{N}^{-1}\,\bd{U}\r)^{-1}
\l(\bd{N}^{-1}\,\bd{U}\r)^t\r]\, \l(\bd{d} - \bd{A}\,\bd{s}\r).
\label{eqn:likeMpixOffMarg}
\end{eqnarray}
Here, $\bd{U}$ is a `mixing matrix' for the map offsets, i.e., a matrix of a rank $(n_f \, n_s \, n_{pix})\,\times\,(n_f \, n_s)$, such as,
\begin{eqnarray}
\bd{U}^t \equiv \l[ \bd{I}_{\l(n_s\,n_f\r) \times \l(n_s\,n_f\r)}, \dots, \bd{I}_{\l(n_s\,n_f\r)\times \l(n_s\,n_f\r)}\r]
\end{eqnarray}
where $\bd{I}_{n\times n}$ is a square unit matrix of a rank $n$,
and $\bd{\Xi} \equiv {\rm diag} \{ \xi_i\}$.
Hereafter, we will refer to the matrix appearing in equation~(\ref{eqn:likeMpixOffMarg}) as $\bd{M_\Xi}^{-1}$. 
Note that its form could have been anticipated, and obtained as a result of the application of the 
Sherman-Morrison-Woodbury formula \citep{Golub_book} to a matrix, $\bd{M_\Xi}$,
\begin{eqnarray}
\bd{M_\Xi} \equiv \bd{N} + \bd{U} \, \Xi \, \bd{U}^t.
\label{eqn:mMatDef}
\end{eqnarray}
The latter matrix is nothing else but the input map noise correlation matrix
with an extra uncertainty due to the random offsets included
\citep[e.g.,][]{stompor_etal_2002}.

The likelihood in equation~(\ref{eqn:likeMpixOffMarg}) has the standard
(generalised) ``$\chi^2$'' form with the covariance given now by the
matrix $\bd{M_\Xi}$.  Consequently, all the formulae derived in the
previous sections could, in principle, be applied straightforwardly to
our case at hand by replacing $\bd{N}^{-1}$ by
$\bd{M_\Xi}^{-1}$. Nevertheless caution needs to be exercised
whenever the matrix $\bd{M_\Xi}^{-1}$ turns out to be singular. For instance,
this is the case whenever no prior knowledge about the
offsets is available, i.e., $\bd{\Xi}^{-1} = 0$. Readily, we then have
(see equation~\ref{eqn:mMatDef}),
\begin{eqnarray}
\bd{M}^{-1}\,\bd{U} = 0,
\label{eqn:Msing}
\end{eqnarray}
and thus $\bd{M}^{-1}$ is singular and has as many singular vectors as the arbitrary offsets. 
(Hereafter, $\bd{M}^{-1}$ denotes the matrix $\bd{M_\Xi}^{-1}$ in the no-prior limit: $\bd{\Xi}^{-1} \rightarrow 0$.)

Note that, although for a nearly full sky experiment one could hope to
have a constraint on the relative offsets, at least in some cases, 
from considerations of the foreground-free, high Galactic
latitude regions of the sky, for the small scale experiments such
constraints will most likely not be available. The no prior case is
therefore of significant, practical importance, and we will consider
it and its consequences on the formalism developed here in the
following sections.

\subsubsection{The spectral parameters in presence of random offsets}

\label{sect:specParsWoff}
 
 From equation~(\ref{eqn:likeMpixOffMarg}) we have in the limit of disappearing prior constraints
 on the offsets, i.e., $\bd{\Xi}^{-1} \rightarrow 0$,
 \begin{eqnarray}
-2\, \ln\,{\cal L}'_{data}\l(\bd{s}, \bd{\beta}\r) = \const + \l(\bd{d}-\bd{A}\,\bd{s}\r)^t\, \bd{M}^{-1}\,\l(\bd{d}-\bd{A}\,\bd{s}\r).
\label{eqn:likeMpixOffMargNoPrior}
 \end{eqnarray}
 The singular vectors of the matrix $\bd{M}^{-1}$ define flat directions of
 the likelihood, i.e., those for which the likelihood value remains the same. 
 For any choice of the spectral parameters these flat directions correspond
 to the changes of the component maps, $\delta\,\bd{s}$, which, when projected
 to the map domain, 
 $\bd{A}\,\delta\,\bd{s}$, result in vectors contained in the null space of $\bd{M}^{-1}$.
 The latter is spanned  by the offset vectors, $\bd{U}_i$, (see equation~\ref{eqn:Msing}). 
 The presence of the flat
 directions in the likelihood however does not necessarily mean that 
 no constraint can be set, first, on the spectral parameters and, later, on the component 
 maps. We can avoid the unconstrained modes by restricting the allowed parameter 
 space so that the only included 
 solutions are those for which $\l(\bd{A}^t \, \bd{s}\r)^t\,\bd{U} = 0$. We then take every such 
 a solution to represent an entire class of solutions, which differ by some $\delta\,\bd{s},$ such
that $\bd{A} \, \delta \, \bd{s}$ is contained within the subspace spanned by the offset
vectors, $\bd{U}$, and as such, given the data model, are indistinguishable.
 We thus modify the likelihood in equation~(\ref{eqn:likeMpixOffMargNoPrior})
 by introducing a penalty function, which forces the likelihood to decay rapidly
 along the flat directions, thus effectively excluding undesired solutions.
The relevant likelihood expression then reads,
\begin{eqnarray}
-2\,\ln\,{\cal L}''_{data}\l(\bd{s}, \bd{\beta}\r)
& \equiv & \const  + \l(\bd{d}-\bd{A}\,\bd{s}\r)^t\, \bd{M}^{-1}\, \l(\bd{d}-\bd{A}\,\bd{s}\r)
  + \gamma^2 \, \l[\l(\bd{A}\,\bd{s}\r)^t\,\bd{U}\r]\,\l[\l(\bd{A}\,\bd{s}\r)^t\,\bd{U}\r]^t \nonumber \\
  & = &  \const + \l(\bd{d}-\bd{A}\,\bd{s}\r)^t\, \bd{M}^{-1}\, \l(\bd{d}-\bd{A}\,\bd{s}\r)
  + \gamma^2 \, \bd{s}^t\,\l(\bd{A}^t\,\bd{U}\r)\,\l(\bd{A}^t\,\bd{U}\r)^t\,\bd{s},
\label{eqn:likeMpixOffMargNoPriorMod}
 \end{eqnarray}
 where $\gamma^2$ is a positive and sufficiently large number.
 We can see that the above likelihood effectively restricts the allowed component maps, $\bd{s}$,
 to those which are orthogonal to the subspace spanned by $\bd{A}^t\,\bd{U}$. 
 A similar approach to ``weighing out'' unwanted contributions to a likelihood has been proposed 
 in other contexts, e.g., power spectrum estimations \citep[][]{BJK_1998}.
 The vectors
 $\bd{A}^t\,\bd{U}$ are not generally orthogonal, but they can always be replaced by their linear
 combinations which are, whenever it is required or convenient. That can be done using any of
 the  standard orthogonalization procedures \citep[e.g.][]{Golub_book}. These orthogonal vectors 
 can be inserted in the formulae presented below, instead of the vectors,  $\bd{A}^t\,\bd{U}$.
 
  We can use equation~(\ref{eqn:likeMpixOffMargNoPriorMod}) to derive the likelihood for the
 spectral parameters, which now reads,
 \begin{eqnarray}
 -2\ln\,{\cal L}''_{spec}\l(\bd{\beta}\r) \equiv \hbox{{\sc const}} 
- \l( \bd{A}^t\,\bd{M}^{-1}\,\bd{d}\r)^t\,\l(\bd{A}^t\,\bd{M}^{-1}\,\bd{A} + \gamma^2\,
 \l(\bd{A}^t \, \bd{U}\r) \, \l(\bd{A}^t \, \bd{U}\r)^t \r)^{-1}
\l( \bd{A}^t\,\bd{M}^{-1}\,\bd{d}\r).
\label{eqn:slopelikeMpixUnbiasedOffMargLimit}
 \end{eqnarray}
 We note here that the matrix $\bd{A}^t\,\bd{M}^{-1}\,\bd{A}$ may be singular given the
 singularity of $\bd{M}^{-1}$, and thus the second term, i.e., 
 $\gamma^2 \, \l(\bd{A}^t \, \bd{U}\r) \, \l(\bd{A}^t \, \bd{U}\r)^t$ has to correct for that
 if the likelihood  in the above equation is to be well-defined.
  In fact, it is straightforward to show that the singular eigenvectors of 
 $\bd{A}^t\,\bd{M}^{-1}\,\bd{A}$ are indeed all contained in the subspace spanned by
 $\bd{A}^t\,\bd{U}$. This follows from the observation that if $\bd{x}$ is a singular
 eigenvector of $\bd{A}^t\,\bd{M}^{-1}\,\bd{A}$, i.e., $\bd{A}^t\,\bd{M}^{-1}\,\bd{A}\,\bd{x} = 0$,
 and the mixing matrix, $\bd{A}$, is non-singular, i.e., $\bd{A}\,\bd{z} = 0 \iff \bd{z} = 0$,
 then $\bd{M}^{-1}\,\l(\bd{A}\,\bd{x}\r) = 0$, and $\bd{A}\,\bd{x}$ is indeed a singular vector of
 $\bd{M}^{-1}$ and thus belongs to its null space spanned by $\bd{A}^t\,\bd{U}$.
 Consequently, the relevant matrix  in equation~(\ref{eqn:slopelikeMpixUnbiasedOffMargLimit}) 
 is invertible as long as $\gamma^2$ is non-zero.
 We also note that in general the singular eigenvectors of $\bd{A}^t\,\bd{M}^{-1}\,\bd{A}$ 
 do not have to span the entire subspace defined by $\bd{A}^t \, \bd{U}$. However, the
 latter is the case whenever a single set of spectral parameters is adopted for all considered
 pixels.
 
\begin{figure} 
\centerline{
\includegraphics[height = 2in, angle = 0]{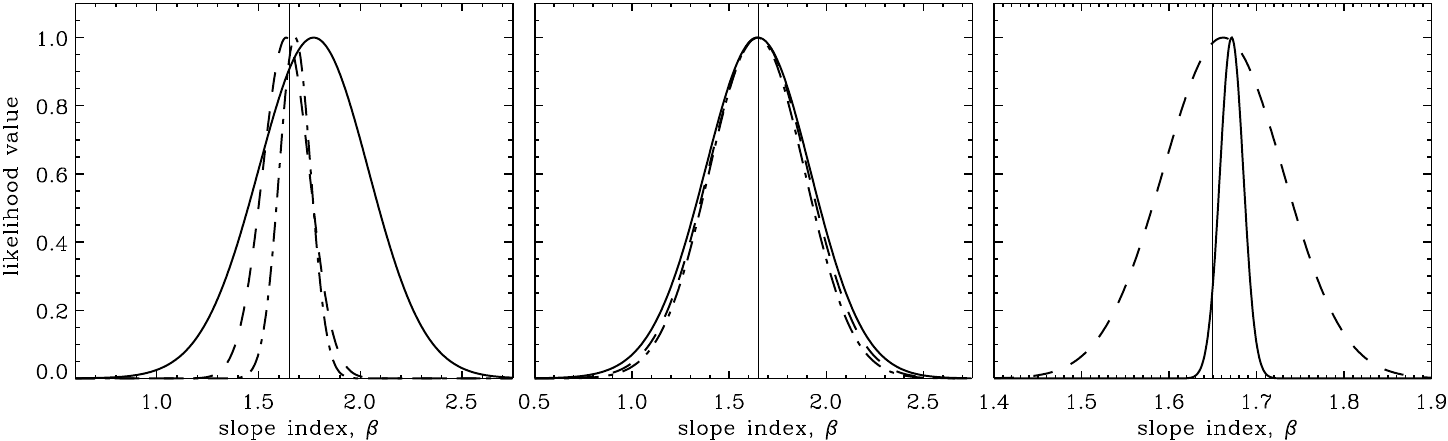}
}
\caption{The one dimensional likelihoods for the slope index with the offsets of
the input maps marginalised over (equations~\ref{eqn:slopelikeMpixUnbiasedOffMargLimit} and
\ref{eqn:specWoffAlgo}). The presented results are for the total intensity data.
{\bf Left panel: } The solid, dashed and dot-dashed lines correspond to the 
cases with $n_{pix} = 100$, $500$ and $1000$ high resolution pixels. 
{\bf Center panel: } The same likelihoods as in the left panel, but this time the results
have been rescaled by a factor of $\sqrt{n_{pix}}$ and recentered at the $\beta = 1.65$
value. This shows that the uncertainty on slope determination decays with a number of
pixels as roughly $\sim 1/\sqrt{n_{pix}}$, once sufficiently many pixels are included.
{\bf Right panel:} A comparison of the two likelihoods computed with (dashed) and
without (solid) marginalisation over the offsets. Both results are for $n_{pix} = 1000$
high resolution pixels and show the loss of precision incurred due to the marginalisation.
The latter depends crucially on the amount of power at the offset scale as present in the input data,
and it is consequently typically found much smaller in the applications to the (simulated) polarized data.
The thin, vertical lines mark the true value of $\beta=1.65$.
\label{fig:offMarg}
}
\end{figure}

 We thus conclude that the likelihood in equation~(\ref{eqn:slopelikeMpixUnbiasedOffMargLimit}) is indeed 
 well-defined for any choice of the
 spectral parameters, what shows that, indeed, the degeneracy of the matrix, $\bd{M}^{-1}$, 
does not lead to any degeneracy in the parameter space of the spectral parameters. 
Nonetheless it does affect the final uncertainty of their determination.
In the case of a single  spectral parameter, slope index, the same for all pixels, this is shown in the left
panel of Figure~\ref{fig:offMarg}. In particular, if only one pixel is observed no limit can be set on the 
slope value \citep[see also,][]{eriksen_etal_2008}. 
For a fixed number of pixels, the loss of a precision with which the slope can be constrained is significant 
as seen in the right panel of Figure~\ref{fig:offMarg}.
The error, however, usually decreases quickly with a number of included pixels and follows a similar 
dependence on a number of the pixels as in a case with the offsets fixed
(see the middle panel of Figure~\ref{fig:offMarg}). Therefore, useful constraints on the spectral parameters 
can be usually derived even in the case of arbitrary offsets of the input maps.

From the numerical perspective, the likelihood calculations can be performed using an efficient iterative 
matrix-vector multipliers similar to those already discussed before. To facilitate that it is useful to rewrite the
expression for the likelihood in equation~(\ref{eqn:slopelikeMpixUnbiasedOffMargLimit}).
In order to do that, we first introduce matrices $\bd{P}$ and $\bd{R}$ defined as,
\begin{eqnarray}
\bd{P} \equiv \l[ \bd{A}^t\, \bd{N}^{-1}\,\bd{U},\ \ \ \bd{A}^t\,\bd{U}\r],
\ \ \ \ \ \ 
\bd{R} \equiv
\l[
\begin{array}{c c}
\ds{-\l(\bd{U}^t\,\bd{N}^{-1}\,\bd{U}\r)^{-1}}  & \ds{\bd{0}} \\
\ds{\bd{0}}                                                  &  \ds{\gamma^2\,\bd{I}_{\l(n_s\,n_f\r)\times \l(n_s\,n_f\r)}}
\end{array}
\r].
\label{eqn:pqrDefs}
\end{eqnarray}
and note that,
\begin{eqnarray}
\bd{A}^t\,\bd{M}^{-1}\,\bd{A} + \gamma^2\, \l(\bd{A}^t\,\bd{U}\r) \, \l(\bd{A}^t\,\bd{U}\r)^t
\equiv 
\bd{A}^t\,\bd{N}^{-1}\,\bd{A} + \bd{P}\,\bd{R}\,\bd{P}^t.
\end{eqnarray}
We then insert the above expression into equation~(\ref{eqn:slopelikeMpixUnbiasedOffMargLimit}) 
and use the Sherman-Morrison-Woodbury formula to obtain finally,
\begin{eqnarray}
\l(\bd{A}^t \, \bd{M}^{-1} \, \bd{d}\r)^t \, \l[\bd{A}^t \, \bd{M}^{-1} \, \bd{A} + \gamma^2 \, \l(\bd{A}^t\,\bd{U}\r)\,
\l(\bd{A}^t\,\bd{U}\r)\r]^{-1}\, \bd{A}^t \, \bd{M}^{-1} \, \bd{d} =
\l(\bd{A}^t \, \bd{M}^{-1} \, \bd{d}\r)^t \, \l(\bd{A}^t \, \bd{N}^{-1} \, \bd{A}\r)^{-1}\, \bd{A}^t \, \bd{M}^{-1} \, \bd{d} + \nonumber \\
-  \l(\bd{A}^t \, \bd{M}^{-1} \, \bd{d}\r)^t \,
\l[ \l(\bd{A}^t \, \bd{N}^{-1} \, \bd{A}\r)^{-1} \, \bd{P}\r] \, \l[ \bd{R}^{-1} + \bd{P}^t \, \l(\bd{A}^t \, \bd{N}^{-1} \, \bd{A}\r)^{-1} \, \bd{P}\r]^{-1}
\,  \l[ \l(\bd{A}^t \, \bd{N}^{-1} \, \bd{A}\r)^{-1} \, \bd{P}\r]^t \, \bd{A}^t \, \bd{M}^{-1} \, \bd{d}.
\label{eqn:specWoffAlgo}
\end{eqnarray}
This expression can be efficiently evaluated performing the matrix vector
multiplications from outside-inwards. Moreover, the likelihood calculation rewritten in this form
can be freed of any ambiguity related to the choice of $\gamma^2$, by simply setting $\gamma^{-2}$
appearing in the expression for $\bd{R}^{-1}$ directly to zero.

We also point out that if we allow for only one set of spectral parameters describing
the frequency scalings of all considered pixels and assume homogeneous 
noise for all input maps, then 
$\l(\bd{A}^t \, \bd{N}^{-1} \, \bd{A}\r)^{-1} \, \bd{P}$ belongs
to the subspace spanned by the vectors $\bd{A}^t\,\bd{U}$
and therefore the
second term on the right hand side of equation~(\ref{eqn:specWoffAlgo})
vanishes.  In this case the calculations of the spectral likelihood
with the offset amplitudes marginalised over is remarkably
simple. This can be done applying formula from
equation~(\ref{eqn:slopelikeMpixUnbiased}) but subtracting the offsets from
the input maps prior to the computation.

\subsubsection{Amplitude determination in the presence of the random offsets and fixed spectral parameters}

\label{sec:ampWoff}

With the spectral parameters fixed we can attempt to determine the
amplitudes of the components, while accounting for the arbitrary
offsets of the input maps. This can be done using the expression analogous
to equation~(\ref{eqn:step2amps}) but derived for the modified likelihood,
equation~(\ref{eqn:likeMpixOffMargNoPriorMod}). This leads to,
\begin{eqnarray}
\bd{s}  =  \l(\bd{A}^t\,\bd{M}^{-1}\,\bd{A} + \gamma^2 \, \l(\bd{A}^t\,\bd{U}\r)\,
\l(\bd{A}^t\,\bd{U}\r)^t\r)^{-1}\,\bd{A}^t\,\bd{M}^{-1}\,\bd{d},
\label{eqn:step2ampsWoff}
\end{eqnarray}
where $\gamma^2$ is the same as before. Using the results derived in the previous
Section (equation~\ref{eqn:specWoffAlgo}) we can rewrite the above equation as,
\begin{eqnarray}
\bd{s} & = &
\l(\bd{A}^t \, \bd{N}^{-1} \, \bd{A}\r)^{-1}\, \bd{A}^t \, \bd{M}^{-1} \, \bd{d} 
\label{eqn:step2ampsWoffAlgo}
\\
&-&  \l(\bd{A}^t \, \bd{M}^{-1} \, \bd{d}\r)^t \,
\l[ \l(\bd{A}^t \, \bd{N}^{-1} \, \bd{A}\r)^{-1} \, \bd{P}\r] \, \l[ \bd{R}_0^{-1} + \bd{P}^t \, \l(\bd{A}^t \, \bd{N}^{-1} \, \bd{A}\r)^{-1} \, \bd{P}\r]^{-1}
\,  \l[ \l(\bd{A}^t \, \bd{N}^{-1} \, \bd{A}\r)^{-1} \, \bd{P}\r]^t \, \bd{A}^t \, \bd{M}^{-1} \, \bd{d},
\nonumber
\end{eqnarray}
where the matrices $\bd{P}$ and $\bd{R}$ are defined in equation~(\ref{eqn:pqrDefs}) and 
$\bd{R}_0^{-1} = \bd{R}\l(\gamma^{-2} = 0\r)$, and
thus the above equation is strictly independent on a value of $\gamma^2$.

As it is clear from our previous discussion the actual solution is not unique and any
other, equivalent solution can be obtained by adding to equation~(\ref{eqn:step2ampsWoff})
any vector from the subspace spanned by $\bd{A}^t\,\bd{U}$.
These unconstrained modes in the recovered component maps depend on the mixing 
matrix as well as best-fitting spectral parameters found on the first step of the
procedure.  For example, in a case of the matrix $\bd{A}$ being
pixel-independent discussed already above they will be likewise pixel independent
and will result in the total offset of all the derived component maps being
unconstrained. For spatially-dependent frequency scalings the unconstrained modes 
will be more complex, potentially giving rise to some degree of arbitrariness in the spatial pattern 
of the recovered signals and therefore may have more pronounced consequences.
The important fact is that the arbitrary modes can always be
straightforwardly determined
and therefore the relevant information passed to the next stages
of the map processing such as power spectrum estimation, and the loss
of precision due to their presence can
be properly included in the estimated errors of the entire analysis
chain \citep[e.g.,][]{BJK_1998}.  

It is interesting to note that the problem of the arbitrary offsets considered here
is just a generalization of a simpler case involving combining 
multiple maps at the same frequency, which may
completely or only partially overlap on the sky
\citep[][]{stompor_etal_2002}. In such a case the constraint on
relative offsets between any pair of the maps can be always derived if
only the two maps overlap. However, their absolute offsets remain
arbitrary.

As in Section~\ref{sect:compAmpDet} we can use the curvature matrix formalism to describe
the covariance pattern of the recovered components. The relevant formulae
are listed in Appendix~\ref{app:curvFullLike} (equation~\ref{eqn:noiseCorrFromCurvMatWoff}).
We stress that those equations do not incorporate the unconstrained modes,
which need to be accounted for separately.

From the point of view of a numerical implementations, the same efficient techniques as discussed
in the previous section can be applied in the computations required here.

The effects of the unknown offsets are also discussed in \citet{eriksen_etal_2008}. Those authors have
introduced a prior to break the degeneracy between the sky parameters and the offsets in the data likelihood 
in equation~(\ref{eqn:genMpixLikeOff}) in order to estimate the offset values.
This is in contrast with the approach proposed here, which marginalises over the offsets and then incorporates 
the additional uncertainty on the subsequent stages of the separation process.

\subsection{Calibration errors}

\label{sect:calib}

The overall calibration of the single frequency maps is usually known only with 
some non-zero precision which for the small scale experiments can be as large
as $5-10\%$ of the best-fitting value. It is therefore important to account for the calibration
uncertainty in the likelihood considerations of the sort presented here.

We will denote hereafter the calibration factors at different frequencies and Stokes parameters, 
as $\bd{\omega}_i$. As offsets, these coefficients are assumed to be pixel-independent as they 
characterise the map as a whole. We can then write the initial likelihood as (cf. equation~\ref{eqn:genMpixLike}),
\begin{eqnarray}
-2\,\ln\,{\cal L}_{data}\l(\bd{s}, \bd{\omega}, \bd{\beta}\r) =  \const + 
\l(\bd{d}-\bd{\Omega}\,\bd{A}\,\bd{s}\r)^t\,\bd{N}^{-1}\l(\bd{d}-\bd{\Omega}\,\bd{A}\,\bd{s}\r).
\label{eqn:genLikeMpixCalib}
\end{eqnarray}
Here the noise correlation matrix $\bd{N}^{-1}$ is assumed not to contain any
correction due to the arbitrary offsets. Next we will consider the complete case. 
The calibration matrix, $\bd{\Omega}$, is then block-diagonal with each block, 
$\bd{\Omega}_p$, corresponding to a different sky pixel, being pixel independent, 
diagonal, and given by,
\begin{eqnarray}
\bd{\Omega}_p \equiv {\rm diag} \l\{\bd{\omega}_i\r\}_{i=0,\dots,n_f\,n_s-1}.
\end{eqnarray}
Alternately, we also arrange all the calibration coefficients in a
vector and refer to it as, $\bd{\omega}$.

In general, the two matrices, $\bd{\Omega}$ and $\bd{A}$, do not
commute and therefore the calibration errors of the single frequency
maps can not be straightforwardly incorporated as the calibration
error of the components. However, we can always rescale all the
components by some arbitrary factor at the same time rescaling the
calibration matrix, $\bd{\Omega}$. This clearly gives rise to a
presence of the degeneracy in the likelihood problem, which can be
broken if the prior constraints are available.  Consequently, in the
following, we will assume that the calibration factors have been
measured with some accuracy and their error distribution is well
represented by Gaussian with a error matrix, $\bd{\Sigma}$.  In the
typical case, all $\bd{\omega}_i$ are assumed to be measured
independently with an RMS precision, $\sigma_{\bd{\omega}_i}^2$,
around some central value, ${\bd{\bar \omega}}_i$. Then,
\begin{eqnarray}
\bd{\Sigma} \equiv {\rm diag}\,\l( {\sigma_{\bd{\omega}_i}^2}\r)_{i=0,n_f-1}.
\label{eqn:sigmaCdiag}
\end{eqnarray}
In a case when the calibration errors are correlated, e.g., due to the
uncertainty on the flux of the common source used for this purpose,
$\bd{\Sigma}$ may be non-diagonal. For example in a case when a single
broad frequency source is used for the calibration of all channels,
the relevant error matrix could read,
\begin{eqnarray}
\bd{\Sigma} = \sigma^2_{source} \l[
\begin{array}{c}
1\\
\vdots\\
1
\end{array}
\r]
\l[
\begin{array}{ c c c}
1 & \dots & 1
\end{array}
\r]
+
\hbox{\rm diag}\,\l({\sigma_{\bd{\omega}_i}^2}\r),
\label{eqn:sigmaCcorr}
 \end{eqnarray}
where the first term is due to the error in our knowledge of the brightness of the calibration source,
and the second due to the instrumental noise for each frequency channel.  
 In addition we will postulate that these
(absolute) errors are Gaussian, and use $\bd{\bar\Omega}$ (or $\bd{\bar\omega}$ for the vectors)
to denote the central values of the calibration parameters. We also posit hereafter that $\bd{\Sigma}$ is 
invertible. 
The first term corresponds to a global calibration error and it will propagate without any changes all the
way to the components amplitudes.
Consequently, in the examples discussed in the following, we will discuss only the 
second term describing the uncorrelated, frequency channel specific errors. We emphasise though that
the formalism presented here is fully general.

Here, we adopt an approach in which we treat the calibration parameters as part of the parametrization 
of the mixing matrix. In this case, the considerations from Section~\ref{sect:specParams} are straightforwardly
applicable. In particular, the spectral likelihood estimator, equation~(\ref{eqn:slopelikeMpixUnbiased}), 
reads in the present case as,
\begin{eqnarray}
-2\,\ln\,{\cal L}_{spec}\l(\bd{\omega}, \bd{\beta}\r) \equiv \hbox{{\sc const}} 
 -  \l( \bd{A}^t\,\bd{\Omega}^t\,\bd{N}^{-1}\,\bd{d}\r)^t\,\l(\bd{A}^t\,\bd{\Omega}^t\,\bd{N}^{-1}\,\bd{\Omega}\,\bd{A}\r)^{-1}
\l( \bd{A}^t\,\bd{\Omega}^t\,\bd{N}^{-1}\,\bd{d}\r) 
+  \l(\bd{\omega}-\bd{\bar{\omega}}\r)^t\,\bd{\Sigma}^{-1}\,\l(\bd{\omega}-
\bd{\bar \omega}\r),
\label{eqn:slopelikeMpixUnbiasedWithCalib}
\end{eqnarray}
where the last term is the prior constraints on the calibration. Note that the likelihood above is not degenerate only if
some prior information is included. Using equation~(\ref{eqn:slopelikeMpixUnbiasedWithCalib}) we can
find the maximum likelihood estimates of the spectral parameters and calibration coefficients and use those
in the component amplitude estimation on the second step of the process. The relevant equation is identical to
equation~(\ref{eqn:step2amps}), but with the mixing matrix $\bd{A}$ replaced by a product of the calibration and mixing matrices, 
$\bd{\Omega}\,\bd{A}$. We can also employ the curvature matrix approximation to estimate the extra uncertainty related to 
the calibration uncertainty.  

\begin{figure}
\centerline{
\includegraphics[height = 1.9in, angle = 0]{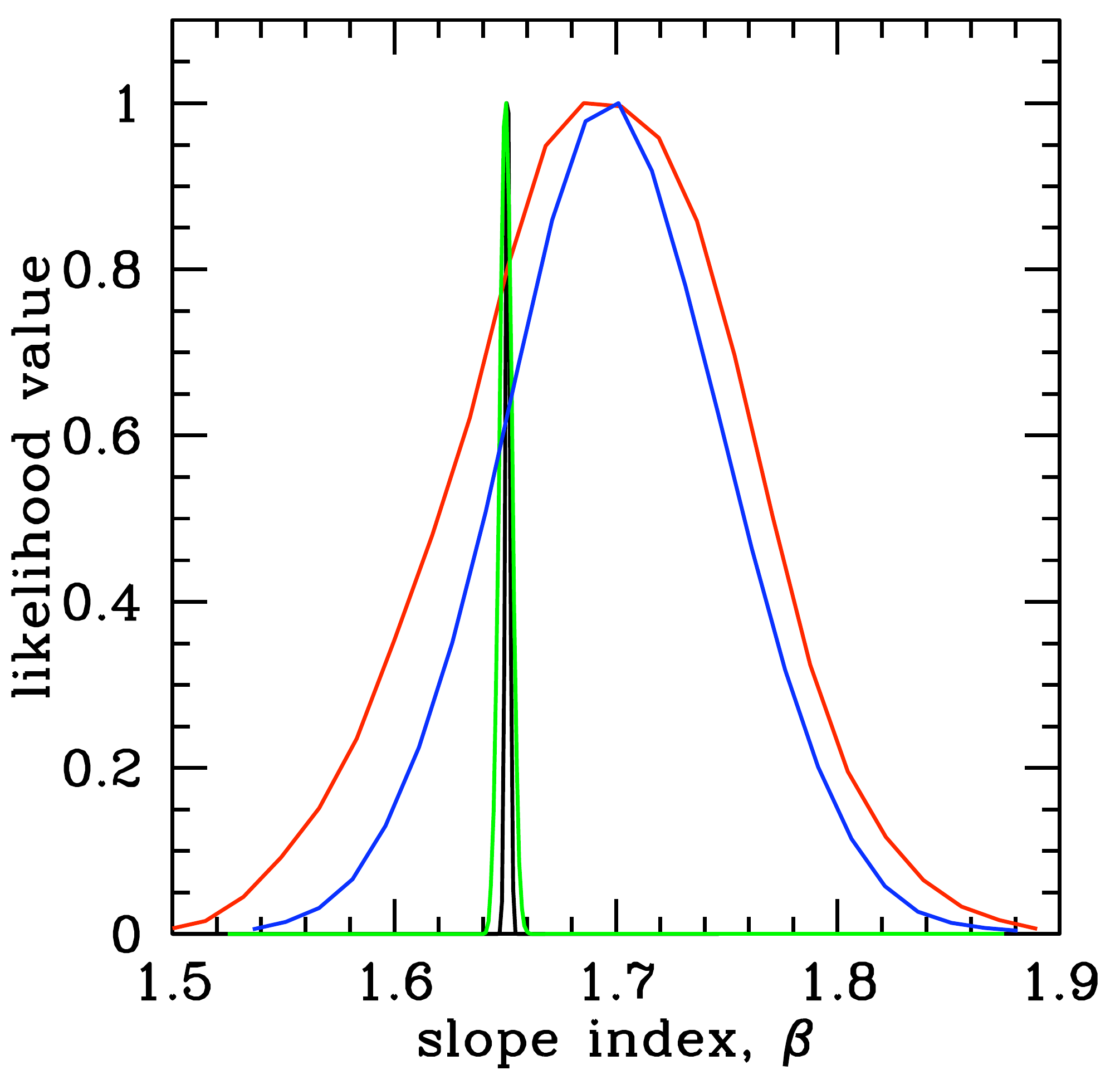}
\includegraphics[height = 1.9in, angle = 0]{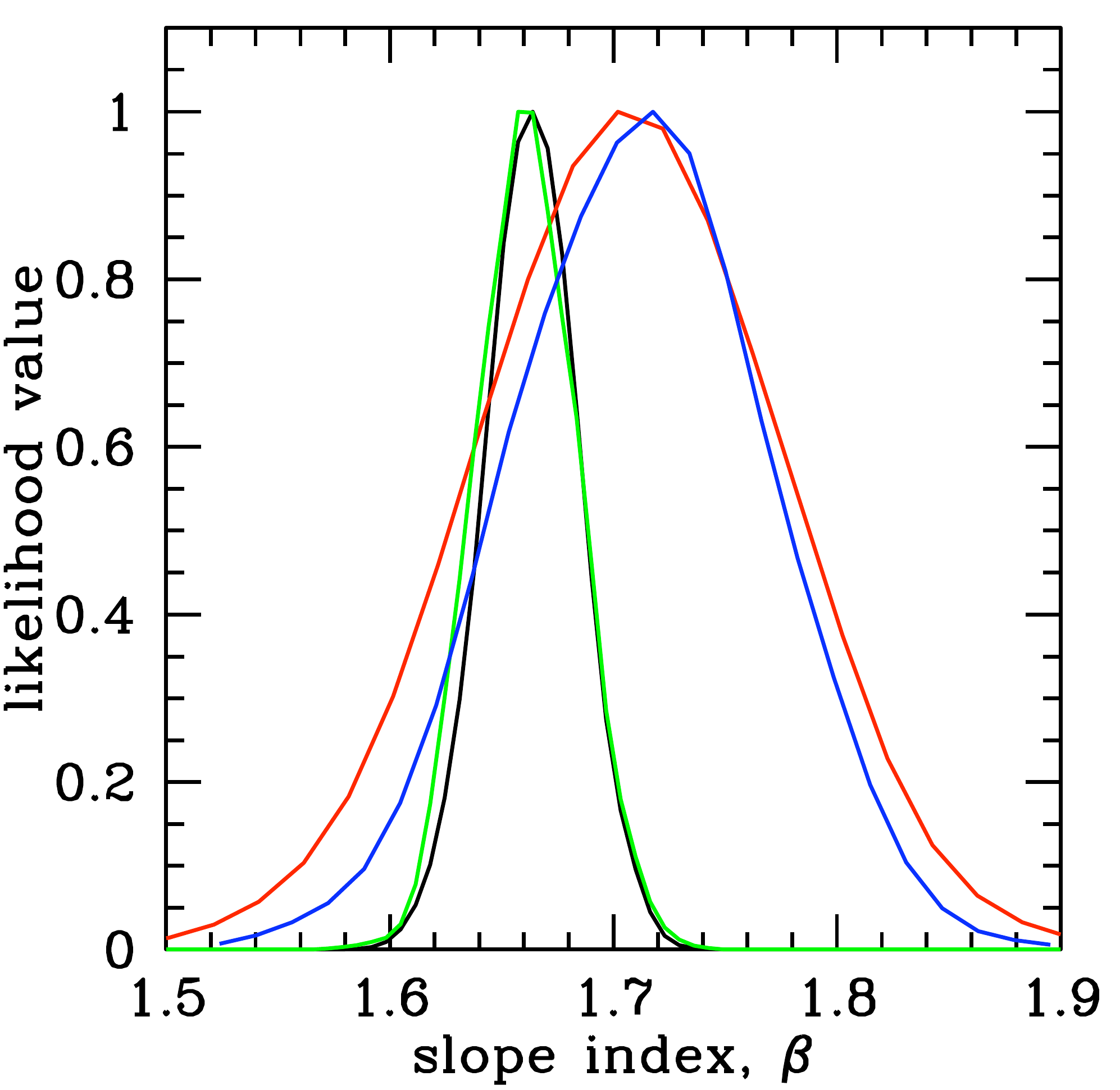}
\includegraphics[height = 1.85in, angle = 0]{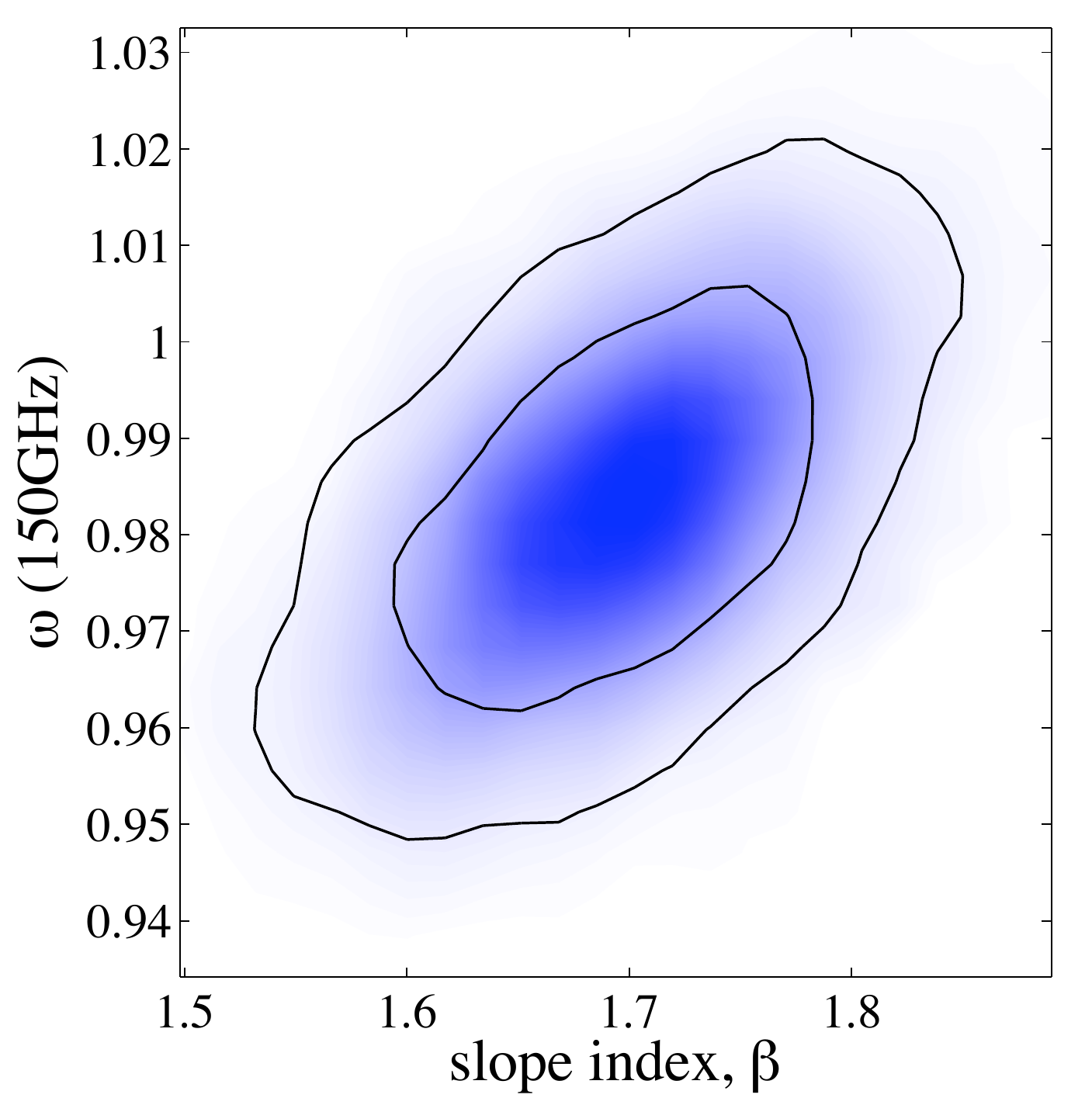}
}
\caption{
The constraints on the slope index $\beta$ derived for four different sets of assumptions
and total intensity and Stokes $Q$ parameter ({\bf left and middle panels}).
These correspond to (from the most stringent to the most lax constraints): (1) the calibration parameters
and map offsets are known perfectly, (2) the calibration parameters are known, the offsets are marginalised
over, (3) the calibration parameters for two high frequency channels are known only with $2$\% precision,
the remaining channel calibration is known, the offsets marginalised over; (4) the calibration parameters for
all three channels are all known with $2$\% accuracy and the offsets are marginalised over.
{\bf Right panel:} the two-dimensional marginalised likelihood of the slope index and the calibration parameter 
for the lowest (150GHz) frequency channel calculated in the case (4).
\label{fig:calibration}}
\end{figure}

\subsection{Combining the two: unknown offsets plus calibration}

We present here our final algorithm for the component separation problem. The algorithm looks for
the maximum likelihood estimates of the spectral parameters, calibration factors, component maps,
while allows for a marginalisation over the arbitrary input maps offsets. The likelihood maximization
is done in two steps. First, we determine the spectral parameters and calibration factors. This is done
by maximization (direct or via MCMC sampling) of the following spectral likelihood 
(this likelihood expression is a straightforward result of combining together 
equations~(\ref{eqn:slopelikeMpixUnbiasedOffMargLimit}) and (\ref{eqn:slopelikeMpixUnbiasedWithCalib})),
\begin{eqnarray}
(\bullet)\ \ -2\,\ln\,{\cal L}_{spec} \l(\bd{\omega}, \bd{\beta}\r) = & - & \l(\bd{A}^t\,\bd{\Omega}^t\,\bd{M}^{-1}\,\bd{d}\r)^t\, \l(\bd{A}^t\,\bd{\Omega}^t\,
\bd{M}^{-1}\,\bd{\Omega}\,\bd{A} + \gamma^2 \, \l(\bd{A}^t\,\bd{\Omega}^t\,\bd{U}\r)\,\l(\bd{A}^t\,\bd{\Omega}^t\,\bd{U}\r)^t\r)^{-1}\,\bd{A}^t\,\bd{\Omega}^t\,\bd{M}^{-1}\,\bd{d} 
\label{eqn:sloplikeMpixFinal}
\\
& + & \l(\bd{\omega}-\bd{\bar{\omega}}\r)^t\,\bd{\Sigma}^{-1}\,\l(\bd{\omega}-\bd{\bar{\omega}}\r) + \hbox{\sc const}.
\nonumber
\end{eqnarray}
On the second step, we estimate the component amplitudes by solving,
\begin{eqnarray}
(\bullet\bullet)\ \ \bd{s} = \l(\bd{A}^t\,\bd{\Omega}^t\,\bd{M}^{-1} \,\bd{\Omega}\,\bd{A}
+\gamma^{2}\,\l(\bd{A}^t\,\bd{\Omega}^t\,\bd{U}\r)\,\l(\bd{A}^t\,\bd{\Omega}^t\,\bd{U}\r)^t\r)^{-1}
\,\bd{A}^t\,\bd{\Omega}^t\,\bd{M}^{-1}\,\bd{d}.
\label{eqn:ampsFinal}
\end{eqnarray}
(c.f., equation~\ref{eqn:step2ampsWoff})
The constant $\gamma^{2}$ is a sufficiently large, positive number, a value of which should be such that the matrix inversion required in the above equations is stable.
The matrix, $\bd{M}^{-1}$, is defined in equation~(\ref{eqn:likeMpixOffMarg}) (with $\bd{\Xi}^{-1} = 0$).

The effects due to the calibration uncertainty in the cases with and without the offsets marginalisation, are
illustrated in Figure~\ref{fig:calibration}. Clearly, its impact on the precision of the determination of the spectral parameters is significant, 
as demonstrated by the substantial broadening of the posterior likelihoods for the slope index in the left panels of 
Figure~\ref{fig:calibration}. This effect exceeds by a large factor the loss of precision due the unknown offsets.
The loss of accuracy incurred in this case is due to the near degeneracy between the calibration factors and the 
slope index as displayed in the right panel of Figure~\ref{fig:calibration} and which effectively is only broken by the 
prior constraints used for the calibration. We stress that the assumed here $2$\% calibration uncertainty 
is quite optimistic in the context of the current small-scale CMB experimentation.

We would like to minimise the effects due to the nuisance factors such as calibration and offset uncertainties.
For that we would have to strive to use complete (or nearly so) maps as an input data vector $\bd{d}$ in equation~(\ref{eqn:sloplikeMpixFinal}). 
That clearly can be done only if general mixing matrices $\bd{A}$, permitting multiple sets of spectral
parameters are used and simultaneously estimated. As we have hinted at earlier, the formalism presented
here is sufficiently general to accommodate such cases.
A more important limitation here is more likely to arise due to the limitation in available
computing resources. One may need therefore to develop strategies as to how to introduce the 
pixel subsets with different spectral parameter set and how large fraction of all pixels has to be used 
in the estimation process not to be significantly affected by our ignorance of the calibrations and offsets. 

\section{Conclusions}

\label{sect:concl}

We have presented a parametric, maximum likelihood algorithm for
separating the components of the sky maps produced by CMB
experiments. The likelihood maximization is performed in two steps
\citep[c.f.,][]{eriksen_etal_2006}: on the first step, we perform a
maximization of the spectral likelihood as given by formula
equation~(\ref{eqn:sloplikeMpixFinal}); on the second step, we use the
derived parameters to estimate the component maps via
equation~(\ref{eqn:ampsFinal}). We have shown that this procedure indeed
leads to the maximum likelihood estimates derived directly from the
full data likelihood (equation~\ref{eqn:genMpixLike}). We worked out
example applications of the algorithm, considering a modeled portion
of the sky containing a mixture of CMB and Galactic dust foreground,
in three frequency channels at 150, 250, 410 GHz, in total intensity
and polarization.

The likelihood approach, described here, is applicable as long as a
suitable data model (e.g., equation~(\ref{eqn:dataModelWoff})) and signal
parametrization is available for each pixel. From the experience of
the data analysis carried out for the past CMB experiments so far, as
well as the current knowledge of the gross features of diffuse
foregrounds, these two requirements are likely to be
fulfilled. The approach is flexible and permits an inclusion of a variety of
priors, in particular those constraining the allowed values of to the spectral 
parameters, as well a number of extensions accounting for the specificity 
of the input data. Any spatial templates for component amplitudes
can be incorporated as additional data sets. Therefore, the approach promises to be
useful for the analysis of the real data sets as those anticipated
from satellite missions \citep[see e.g.,][]{eriksen_etal_2006} and in
particular the balloon-- and ground-- based experiments. In fact we
have shown how to amend the basic algorithm to account for the input
map offset uncertainty and to deal with the calibration error.  
We have found that though both these effects may affect the 
precision of the spectral index recovery, it is the calibration uncertainty,
which, for realistic prior constraints, is likely to be  a dominant factor.
The arbitrary offsets lead in addition to the presence of the unconstrained 
modes in the component maps.
In a simple case of spatially independent mixing matrices these
will only result in the free offsets of the recovered component maps.
However, for more complex mixing matrices these may correspond to the
more complicated modes in each of the component maps. 

\begin{figure} 
\centerline{
\includegraphics[height = 1.85in, angle = 0]{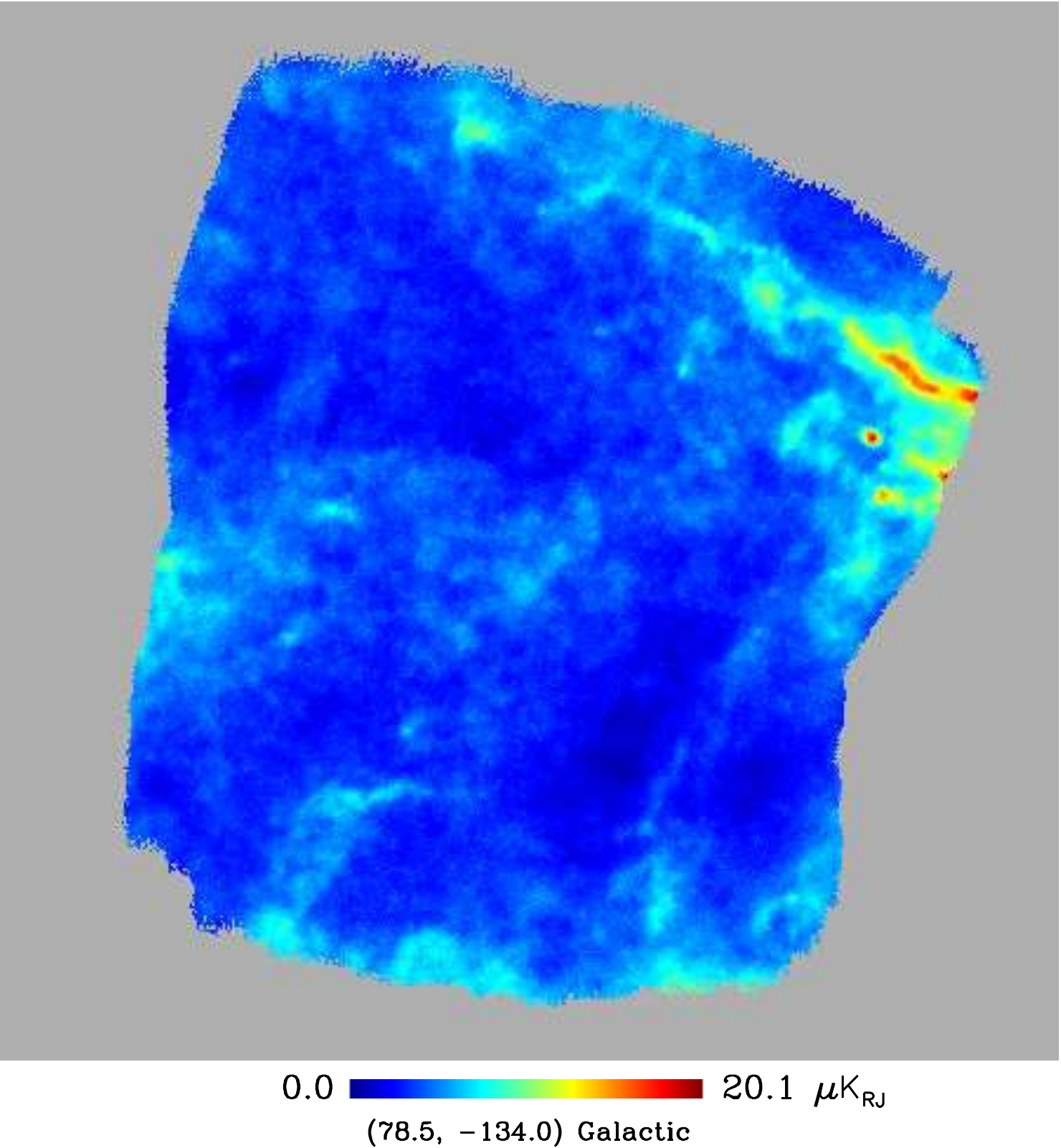}
\includegraphics[height = 1.85in, angle = 0]{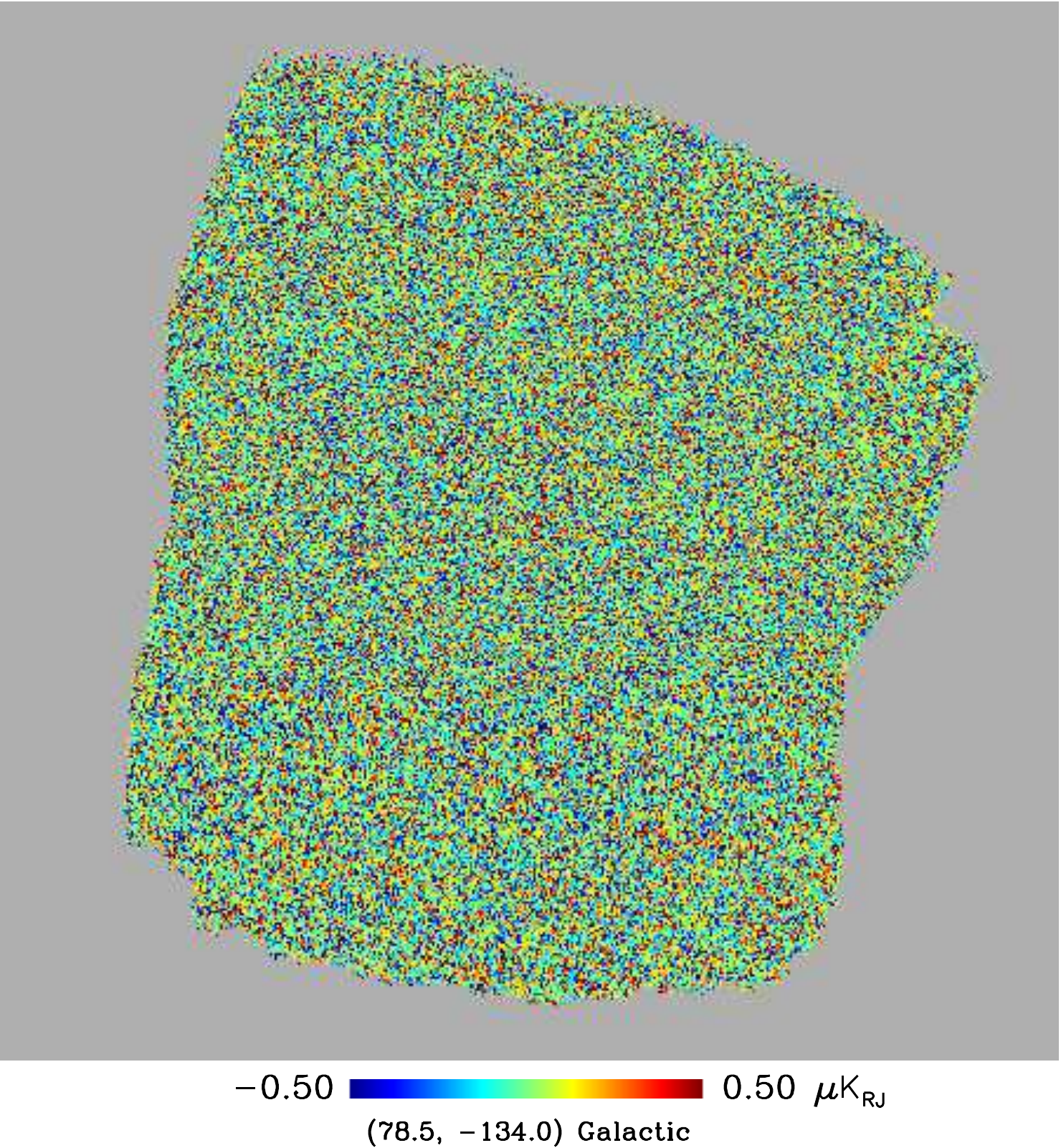}
\includegraphics[height = 1.85in, angle = 0]{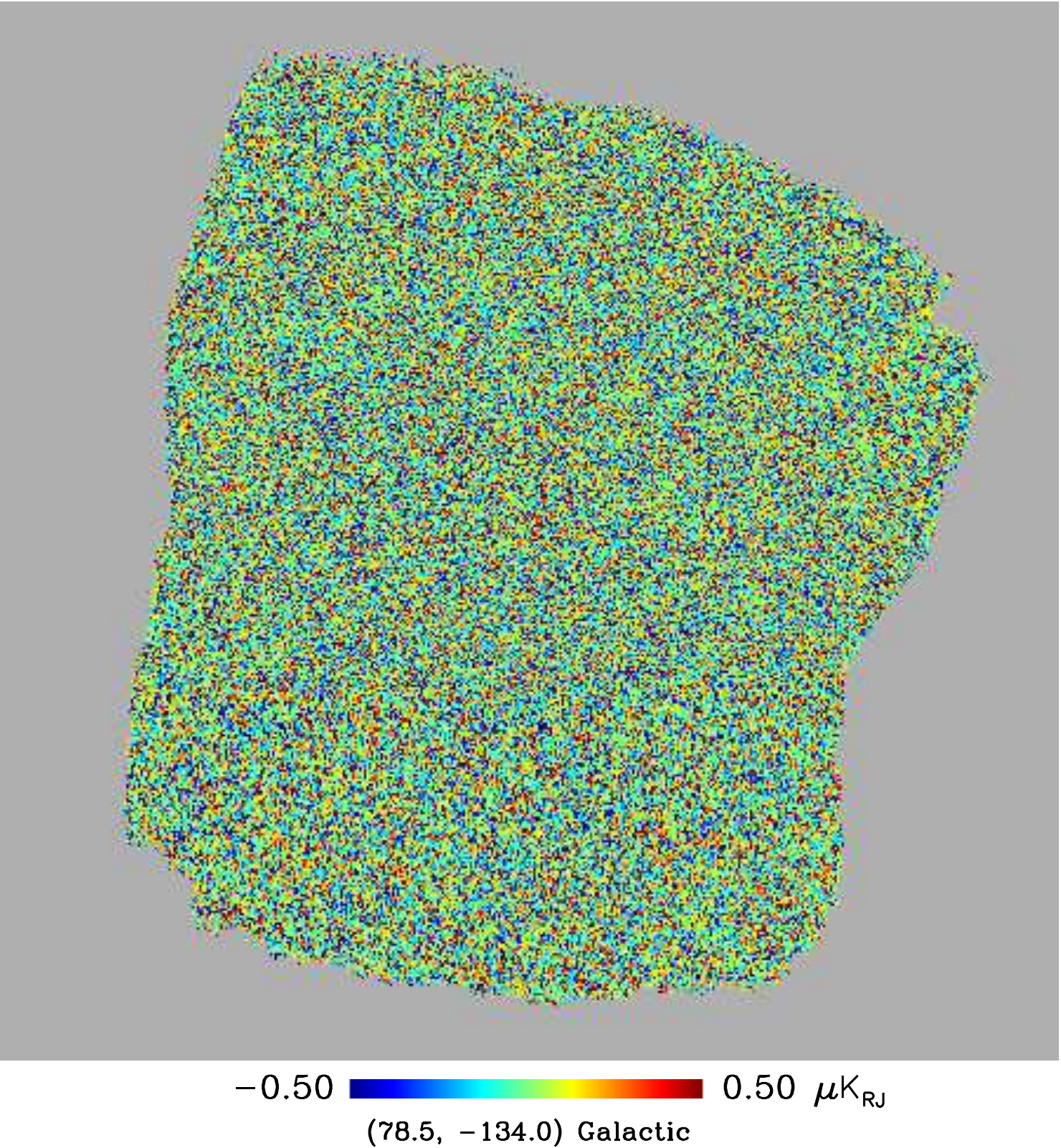}
\includegraphics[height = 1.85in, angle = 0]{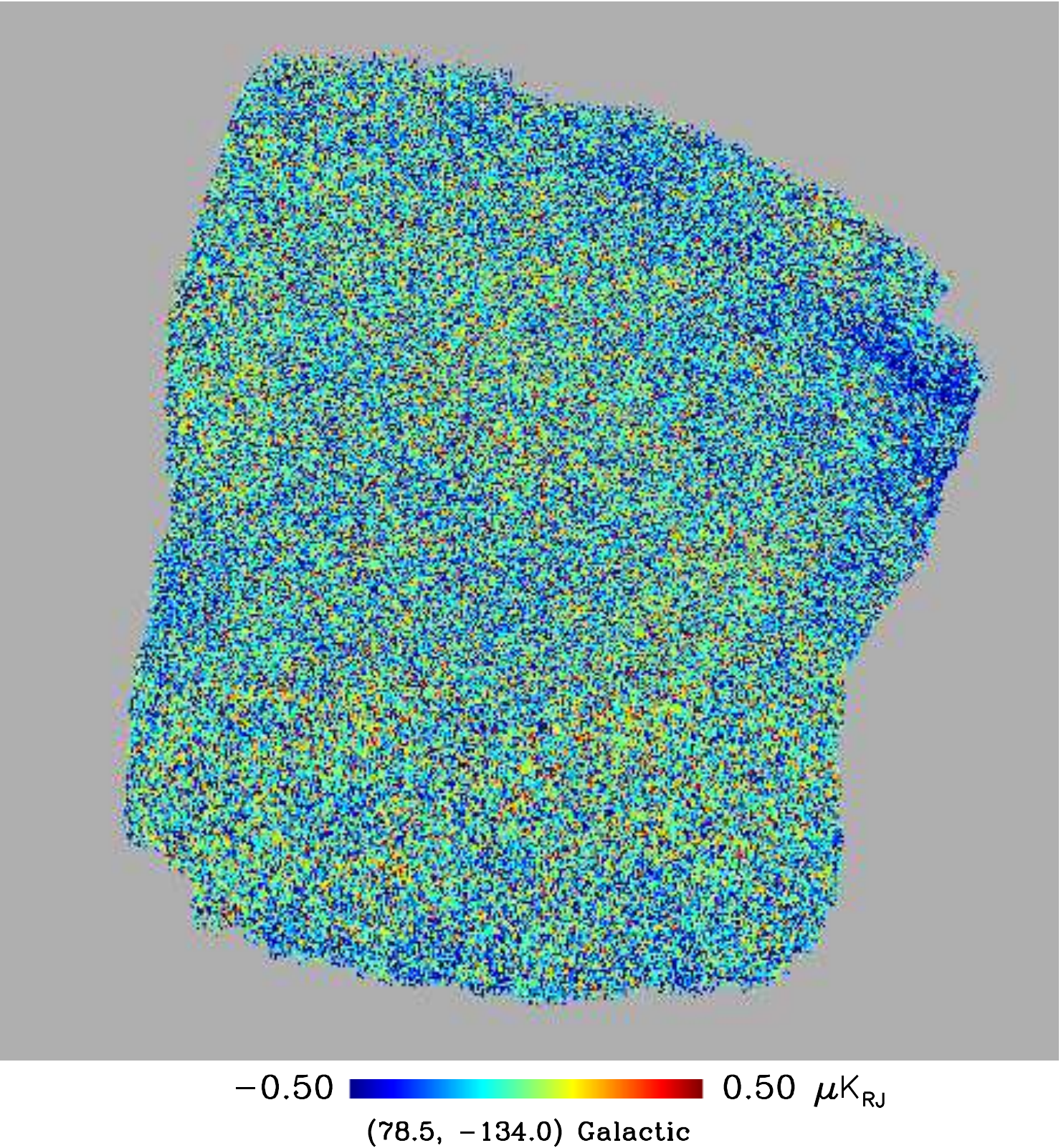}
}
\caption{
The recovered total intensity dust signal ({\bf left panel}) calculated when allowing for the 
calibration uncertainty of $2$\% in all three frequency channels (i.e., the case (3) below). 
The {\bf three following
panels} show the maps of the recovered dust residual and correspond to three specific
test cases in which it is assumed that: (1) the actual values of spectral index and calibration parameters
are known a priori, (2) the calibration parameters are known, but the spectral index is determined from
the data, (3) the calibration parameters are known only with a $2$\% error for each of the three frequency channels,
and therefore their actual values, as well as that of the spectral index, are derived
from the data.
The RMS of the residual maps is found to be: $0.263\mu$K$_{\rm RJ},$
$0.263\mu$K$_{\rm RJ},$ and $0.260\mu$K$_{\rm RJ},$ respectively. The slope index values used in these three cases
are: $\beta = 1.65, \ 1.651$ and $1.685$.
\label{fig:dustResiduals}}
\end{figure}

The examples of the end-to-end runs of the algorithm described here are
shown in Figure~\ref{fig:dustResiduals}. We show there a recovered total intensity
map of the dust component and the maps of the residuals for three different
sets of progressively more relaxed priors. In particular, we observe that in the 
rightmost map of the residuals the actual dust signal is lurking within the noise --
an effect absent in the two other maps shown here.
This is yet another demonstration of the loss of the precision of the spectral 
index determination due to assumed calibration uncertainties. We note that this
effect is apparent only due to a very high signal-to-noise ratio for the total intensity
maps considered here and would not be seen in the corresponding polarized maps
with their lower sky signal and higher noise.
We also note that RMS values of the map residuals are in  good agreement with 
the values derived from the curvature matrix formalism, applied to the
proposed here likelihood functions.

We note that it is possible to treat the method described here 
just as an algorithm for an efficient maximization of the full data likelihood, 
equation~(\ref{eqn:genMpixLike}), avoiding therefore a need of interpreting the 
expression in equation~(\ref{eqn:sloplikeMpixFinal}) as a common likelihood. 
We have however shown in this paper that adopting the latter point of view is likely
to be insightful and useful in the actual CMB data analysis practice, and 
does not affect the method results.

The implementation of the formalism presented in this paper has been
pixel-based, however, we have pointed out that the algorithm can be
straightforwardly and efficiently extended to operate on the time
domain data streams, simplifying the task of the proper error estimate
propagation between the map-making to the component separation stages.

In future works, we will specialise the present algorithm in the
context of the forthcoming CMB experiments.

\section*{acknowledgements}

We would like to thank Jean-Fran{\c c}ois Cardoso for a useful discussion on the `profile'
likelihoods. We also thank him, Jacques Delabrouille, Hans-Kristian Eriksen, Andrew
Jaffe, and, the reviewer, Mark Ashdown for numerous comments on the manuscript. 
\\
Some of the results in this paper have been derived using the HEALPix
\citep{gorski_etal_2005} package.  We acknowledge the use
of the CAMB \citep{lewis_etal_2000} and the CosmoMC
\citep{lewis_etal_2002} packages.  RS acknowledges support of the EC
Marie Curie IR Grant (mirg-ct-2006-036614) and CB that of 
the NASA LTSA Grant NNG04CG90G. This research has been supported 
in part by the ASI Planck LFI Activity of Phase E2 contract.

\appendix

\section{Curvature matrices for the considered likelihoods}

\label{app:curvFullLike}

We list here some useful curvature matrices for the data likelihoods as considered in this paper.

We start from the case of the full data likelihood, equation~(\ref{eqn:genMpixLike}), applicable to the most basic
data model with both calibrations and offsets known precisely.
The relevant expression for the curvature matrix, $\bd{\Lambda}_{p \, q} \equiv -\partial^2\,{\cal L}/\partial \, p \, \partial \, q$, is given by,
\begin{eqnarray}
- \,{\partial^2 \, \ln {\cal L}_{data} \over \partial \bd{s}^2} & = & \bd{A}^t\,\bd{N}^{-1}\,\bd{A},
\label{eqn:curvMatFullLike1}
\\
- \,{\partial^2 \, \ln {\cal L}_{data} \over \partial {\bd{\beta}}_i\,\partial \bd{\beta}_j} & = &
\l( \bd{A}_{,\bd{\beta}_i}\,\bd{s}\r)^t\,\bd{N}^{-1}\, \l( \bd{A}_{,\bd{\beta}_j}\,\bd{s}\r) 
- \l(\bd{A}_{,\bd{\beta}_i\, \bd{\beta}_j}\,\bd{s}\r)^t \, \bd{N}^{-1}\,\l(\bd{d}-\bd{A}\,\bd{s}\r),
\label{eqn:curvMatFullLike2}
\\
-  \, {\partial^2 \, \ln {\cal L}_{data} \over \partial \bd{s} \, \partial \bd{\beta}_i} & = &
    \bd{A}^t\, \bd{N}^{-1}\,\bd{A}_{,\bd{\beta}_i} \, \bd{s}
-  \bd{A}_{,\bd{\beta}_i}^t\, \bd{N}^{-1}\,\l(\bd{d}-\bd{A}\,\bd{s}\r).
\label{eqn:curvMatFullLike3}
\end{eqnarray}

We will interpret the inverse of the curvature matrix, computed at the peak of the likelihood, 
as a good approximation to the covariance matrix for the  parameters estimated via 
maximization of the likelihood. Specifically, in this approximation,
the $\bd{s}-\bd{s}$ diagonal block of the inverse curvature matrix, $\bd{N^\Lambda}_{\bd{s}\,\bd{s}}$, 
describes the correlation pattern  of the recovered sky components, while the $\bd{\beta}-\bd{\beta}$ block, 
$\bd{N^\Lambda}_{\bd{\beta}\,\bd{\beta}}$, provides estimates of the
errors, and their correlations, of the recovery of the spectral parameters.
Using equations~(\ref{eqn:curvMatFullLike1}-\ref{eqn:curvMatFullLike3}), and inverting the curvature matrix by partition,
we can write explicitly relevant expressions for both these matrix blocks. We obtain,
\begin{eqnarray}
\bd{ \tilde N^\Lambda}_{\bd{s}\,\bd{s}} = \bd{\hat{N}}
+ \l[\bd{\hat{N}} \, \l( \bd{A}^t\, \bd{N}^{-1}\,\bd{A}_{,\bd{\beta}}\,\bd{s}- \bd{A}_{,\bd{\beta}}^t\, \bd{N}^{-1} \, \l(\bd{d}-\bd{A}\,\bd{s}\r)\r)\r] \, 
\bd{\tilde N^\Lambda}_{\bd{\beta}\,\bd{\beta}} \, 
\l[\bd{\hat{N}} \, \l( \bd{A}^t\, \bd{N}^{-1}\,\bd{A}_{,\bd{\beta}}\,\bd{s}- \bd{A}_{,\bd{\beta}}^t\, \bd{N}^{-1} \, \l(\bd{d}-\bd{A}\,\bd{s}\r)\r)\r]^t
 \label{eqn:noiseCorrFromCurvMat}
\end{eqnarray}
and
\begin{eqnarray}
\bd{\tilde N^\Lambda}_{\bd{\beta}\,\bd{\beta}} & = & \l\{
\l( \bd{A}_{,\bd{\beta}_i}\,\bd{s}\r)^t\,\bd{N}^{-1}\, \l( \bd{A}_{,\bd{\beta}_j}\,\bd{s}\r) 
- \l(\bd{A}_{,\bd{\beta}_i\, \bd{\beta}_j}\,\bd{s}\r)^t\,\bd{N}^{-1}\,\l(\bd{d}-\bd{A}\,\bd{s}\r)\r.
\label{eqn:sRMS}
\\
& - & \l. 
\l[ \bd{A}^t\, \bd{N}^{-1}\,\bd{A}_{,\bd{\beta}}\,\bd{s} - \bd{A}_{,\bd{\beta}}^t\, \bd{N}^{-1}\, \l( \bd{d}-\bd{A}\,\bd{s}\r)\r]^t
\, \bd{\hat{N}}\,
\l[ \bd{A}^t\, \bd{N}^{-1}\,\bd{A}_{,\bd{\beta}}\,\bd{s} - \bd{A}_{,\bd{\beta}}^t\, \bd{N}^{-1}\, \l( \bd{d}-\bd{A}\,\bd{s}\r)\r]
\r\}^{-1}.
\nonumber
\end{eqnarray}
Here, $\bd{\hat{N}}$ is a `standard' uncertainty
estimate neglecting the error due to the spectral parameters defined in equation~(\ref{eqn:noiseCorrOptim})
and we treat all the matrices $\bd{A}_{,\bd{\beta}}$ and $\bd{A}_{,\bd{\beta}\bd{\beta}}$ as respectively 3 and
4 dimensional with the extra dimensions having a size given by a number of spectral parameters 
to be determined.
The matrix elements are evaluated at the maximum likelihood values of the parameters.
We point out that the correction terms included in equations~(\ref{eqn:noiseCorrFromCurvMat}) and (\ref{eqn:sRMS})
involve the `map-making' type calculations (which may need to be performed on order of $n_{\beta}\ (\ll n_{pix})$
times) and thus can be efficiently computed using the relevant iterative map-making solvers.

Similar expressions for the curvature matrix can be straightforwardly derived also in the case with calibration 
errors included (Section~\ref{sect:calib}). In fact, if the calibration coefficients, $\bd{\omega}$, are considered 
a subset of all 
the spectral  parameters $\bd{\beta}$, the curvature matrix, $\bd{\Lambda}'$, obtained in this case mostly coincides with the matrix $\bd{\Lambda}$, as discussed above, but
an extra term, which needs to be added to its $\bd{\omega}-\bd{\omega}$ block to account for the
calibration prior, equation~(\ref{eqn:genLikeMpixCalib}), i.e.,
\begin{eqnarray}
\bd{\Lambda}'_{\bd{\omega}\,\bd{\omega}} & = & \bd{\Lambda}_{\bd{\omega}\,\bd{\omega}} + \bd{\Sigma}^{-1},\\
\bd{\Lambda}'_{\bd{i}\,\bd{j}} & = & \bd{\Lambda}_{\bd{i}\,\bd{j}}, \ \ \ \ \ \ \ \ \ \ \  \  \hbox{otherwise.}
\end{eqnarray}

The curvature matrix, $\bd{\Lambda}''$, for the case with the input map offsets marginalised over can be 
derived in a similar fashion, but starting off from equation~(\ref{eqn:likeMpixOffMargNoPriorMod}).
First, we get,
\begin{eqnarray}
- {\partial^2 \, \ln {\cal L}''_{data} \over \partial \bd{s}^2} & = & 
\bd{A}^t\,\l(\bd{M}^{-1} + \gamma^2\,\bd{U}\,\bd{U}^t\r) \, \bd{A},
\label{eqn:curvMatFullLikeWoff1}
\\
- {\partial^2 \, \ln {\cal L}''_{data} \over \partial {\bd{\beta}}_i\,\partial \bd{\beta}_j} 
& = &
\l( \bd{A}_{,\bd{\beta}_i}\,\bd{s}\r)^t\,\l(\bd{M}^{-1} + \gamma^2\,\bd{U}\,\bd{U}^t\r)
\, \l( \bd{A}_{,\bd{\beta}_j}\,\bd{s}\r) 
+ \l(\bd{A}_{,\bd{\beta}_i\, \bd{\beta}_j}\, \bd{s}\r)^t \, \bd{M}^{-1} \, \l(\bd{A}\,\bd{s} - \bd{d}\r)
\label{eqn:curvMatFullLikeWoff2}
\\
- {\partial^2 \, \ln {\cal L}''_{data} \over \partial \bd{s} \, \partial \bd{\beta}_i} 
& = &
\bd{A}_{,\bd{\beta}_i}^t\, \bd{M}^{-1} \, \l(\bd{A} \,\bd{s} - \bd{d}\r)
+ \l(\bd{A}_{,\bd{\beta}_i} \, \bd{s}\r)^t\, 
\l(\bd{M}^{-1} + \gamma^2 \, \bd{U}\, \bd{U}^t\r) \, \bd{A},
\label{eqn:curvMatFullLikeWoff3}
\end{eqnarray}
where we have made an explicit use of the fact that $\bd{U}^t\,\bd{A}\,\bd{s} = 0$.
Then we represent the curvature matrix as,
\begin{eqnarray}
{\bd{\Lambda}''} & = & {\bd{\Lambda}''_0} + 
 \l[\bd{A}, \  \bd{A}_{,\bd{\beta}}\,\bd{s}\r]^t\,\l(\bd{M}^{-1} + \gamma^2 \, \bd{U} \, \bd{U}^t\r)
\, \l[\bd{A}, \ \bd{A}_{,\bd{\beta}}\,\bd{s}\r]\nonumber \\
& = & 
{\bd{\Lambda}''_0}  + \bd{P} \, \bd{R} \, \bd{P}^t,
\end{eqnarray}
where we have used the notation from equation~(\ref{eqn:pqrDefs}) with the mixing matrix, $\bd{A}$, replaced by
$ \l[\bd{A}, \  \bd{A}_{,\bd{\beta}}\,\bd{s}\r]$ and defined,
\begin{eqnarray}
\bd{\Lambda}''_0 \equiv
\l[\bd{A}, \  \bd{A}_{,\bd{\beta}}\,\bd{s}\r]^t\,\bd{N}^{-1} \l[\bd{A}, \  \bd{A}_{,\bd{\beta}}\,\bd{s}\r]
+
\l[
\begin{array}{c c}
\ds{0}    & \ds{\bd{A}_{,\bd{\beta}}^t\, \bd{M}^{-1} \, \l(\bd{A} \,\bd{s} - \bd{d}\r)} \\
\ds{\l[\bd{A}_{,\bd{\beta}}^t\, \bd{M}^{-1} \, \l(\bd{A} \,\bd{s} - \bd{d}\r)\r]^t} & 
\ds{\l(\bd{A}_{,\bd{\beta}\, \bd{\beta}}\, \bd{s}\r)^t \, \bd{M}^{-1} \, \l(\bd{A}\,\bd{s} - \bd{d}\r)}
\end{array}
\r].
\end{eqnarray}
Using similar arguments as those already used in the derivation of equation~(\ref{eqn:specWoffAlgo}), 
we first note that the dependence on the $\gamma^2$ factor disappears in the final result for the 
covariance, which can be written down as,
\begin{eqnarray}
\bd{\tilde{N}}^{\bd{\Lambda}''} = {\bd{\Lambda}''_0}^{-1} -
\l({\bd{\Lambda}''_0}^{-1}\,\bd{P}\r)
\l[\bd{R}_0^{-1} + \bd{P}^t\,{\bd{\Lambda''}_0}^{-1}\,\bd{P}\r]
\l({\bd{\Lambda''}_0}^{-1}\,\bd{P}\r)^t.
\label{eqn:noiseCorrFromCurvMatWoff}
\end{eqnarray}
The inverse of the matrix $\bd{\Lambda}_0$ can be computed by partition as before, while the matrix products 
such as $\l({\bd{\Lambda''}_0}^{-1}\,\bd{P}\r)$ can be calculated efficiently using the iterative linear equation solvers.
We also note that the complete estimate of the uncertainty of the recovered component maps needs to take into
account the unconstrained modes as defined by the subspace of the solutions span by the vectors $\bd{A}^t\,\bd{U}$.

\section{Matrix non-diagonality}

\label{app:matOff}

We define here a simple measure of non-diagonality of a square, invertible, diagonal-dominated matrix, 
$\bd{B}$.
  We first split the matrix, $\bd{B}$, into
 diagonal, $\bd{B}_{diag}$ and non-diagonal part, $\bd{B}_{off}$, i.e., 
 \begin{eqnarray}
 \bd{B} \simeq \bd{B}_{diag} + \bd{B}_{off} = \bd{B}_{diag} \l( \bd{1}+ \bd{B}_{diag}^{-1}\,\bd{B}_{off}\r).
 \end{eqnarray}
 If the off-diagonal part is indeed small enough than the inverse of the matrix $\bd{B}$ should be well
 described by,
\begin{eqnarray}
 \bd{B}^{-1} \simeq  \l( \bd{1}- \bd{B}_{diag}^{-1}\,\bd{B}_{off}\r) \, \bd{B}_{diag}^{-1}.
 \end{eqnarray}
 This will be true whenever,
\begin{eqnarray}
\l( \bd{1}- \bd{B}_{diag}^{-1}\,\bd{B}_{off}\r) \, \bd{B}_{diag}^{-1}  \bd{B}_{diag} \l( \bd{1}+ \bd{B}_{diag}^{-1}\,\bd{B}_{off}\r)
= \bd{1} - \l[\bd{B}_{diag}^{-1}\,\bd{B}_{off}\r] \, \l[\bd{B}_{diag}^{-1} \, \bd{B}_{off}\r] \equiv \bd{1} - \bd{\kappa}
\simeq \bd{1},
  \end{eqnarray}
where the matrix, $\bd{\kappa}$ introduced above, is our measure. We posit that the off-diagonal elements of the matrix $\bd{B}$ are negligible, whenever, $\bd{\kappa}_{ij} \simeq 0$, with a precision as required by a problem at hand.

\begin{thebibliography}{99}

\bibitem[Ade\ \etal(2007)]{ade_etal_2007}
Ade, P. \etal, 2007, submitted to \apj.

\bibitem[Ashdown\ et\ al.(2007)]{ashdown_etal_2007}
Ashdown, M.A.J. \etal, 2007, \aa, 467, 761

\bibitem[Barndoff-Nielsen \&\ Cox(1994)]{nielsen_cox_1994}
Barndoff-Nielsen, O.E., \&\ Cox, D.R., 1994, {\it Inference and Asymptotic}, Chapman \&\ Hall, London

\bibitem[Bennett\ \etal(1992)]{bennett_etal_1992}
Bennett, C.L., \etal, 1992, \apjl, 396, L7

\bibitem[Benoit\ et\ al.(2004)]{benoit_etal_2004}
Benoit A. \etal, 2004, \aa, 424, 571

\bibitem[Bischoff\ \etal(2008)]{bischoff_etal_2008}
Bischoff, C. \etal, 2008, submitted to \apj.%, 

\bibitem[Bonaldi\ et\ al.(2006)]{bonaldi_etal_2006}
Bonaldi, A., Bedini, L., Salerno, E., Baccigalupi, C., \&\ de Zotti G., 2006 \mnras, 373, 263 

\bibitem[Bond,\ Jaffe,\ \&\ Knox(1998)]{BJK_1998}
Bond, J.R., Jaffe, A.H, \&\ Knox, L., 1998, \prd, 57, 2117

\bibitem[Brandt\ \etal(1994)]{brandt_etal_1994}
Brandt, W.N., Lawrence, C.R., Readhead, A.C.S., Pakianathan, J.N., \&\ Fiola, T.M., 
1994, \apj, 424, 1

\bibitem[Delabrouille\ \etal(2003)]{delabrouille_etal_2003}
Delabrouille, J., Cardoso, J.-F., \&\ Patanchon, G., 2003, \mnras, 346, 1089

\bibitem[Dunkley\ \etal(2008)]{dunkley_etal_2008}
Dunkley, J., \etal, 2008, submitted to \apjs%, 

\bibitem[Eriksen\ \etal(2006)]{eriksen_etal_2006}
Eriksen, H.K. \etal, 2006, \apj, 641, 665

\bibitem[Eriksen\ \etal(2008)]{eriksen_etal_2008}
Eriksen, H.K., Jewell, J.B., Dickinson, C., Banday, A.J., G\'orski, K.M., \&\ Lawrence, C.R.,
 2008, \apj, 676, 10

\bibitem[Finkbeiner\ \etal(1999)]{finkbeiner_etal_1999}
Finkbeiner, D.P., Davis, M., \&\ Schlegel, D.J., 1999 \apj, 524, 867

\bibitem[Gold\ \etal(2008)]{gold_etal_2008}
Gold, P., \etal, 2008, submitted to \apjs

\bibitem[Golub and van Loan(1996)]{Golub_book}
Golub, G. H., and van Loan, C. F., 1996, Matrix Computations,
The Johns Hopkins University Press, 3rd edition

\bibitem[G\'orski\ \etal(2005)]{gorski_etal_2005}
G\'orski, K.M., Hivon, E., Banday, A.J., Wandelt, B.D., Hansen, F.K., Reinecke, M., 
\&\ Bartelmann, M., 2005, \apj, 622, 759

\bibitem[Hansen\ \etal(2006)]{hansen_etal_2006}
Hansen, F.K., Banday, A.J., Eriksen, H.K., G\'orski, K.M., \&\ Lilje, P.B., 2006, \apj, 648, 784

\bibitem[Jaffe\ \etal(2004)]{jaffe_et_al_2004}
Jaffe, A.H., \etal, 2004, \apj, 615, 55

\bibitem[Johnson\ \etal(2007)]{johnson_etal_2007}
Johnson, B., \etal, 2007, 665, 42

\bibitem[Kamionkowski\ \etal(1997)]{kamionkowski_etal_1997}
Kamionkowski, M., Kosowsky, A., \&\ Stebbins, A., 1997, \prd, 55, 7368

\bibitem[Kogut\ \etal(2007)]{kogut_etal_2007}
Kogut, A. \etal, 2007, \apj, 665, 355

\bibitem[Leach\ \etal(2008)]{leach_etal_2008}
Leach, S.M. \etal, 2008, submitted to \aa

\bibitem[Lewis \&\ Bridle(2002)]{lewis_etal_2002}
Lewis, A., \& Bridle, S., \prd, 66, 103511, 2002

\bibitem[Lewis and Challinor(2006)]{lewis_etal_2006}
Lewis, A., and Challinor, A., \pr, 429, 1, 2006

\bibitem[Lewis, Challinor \&\ Lasenby(2000)]{lewis_etal_2000}
Lewis, A., Challinor, A., \& Lasenby, A., \apj, 538, 473, 2000

\bibitem[Maino\ \etal(2007)]{maino_etal_2007}
Maino, D., Donzelli, S., Banday, A.J., Stivoli, F., \&\ Baccigalupi, C., 2007, 
\mnras, 374, 1207

\bibitem[Masi\ \etal(2006)]{masi_etal_2006}
Masi, S. \etal, 2006, \apj, 458, 687

\bibitem[Montroy\ \etal(2006)]{montroy_etal_2006}
Montroy, T.E. \etal, 2006, \apj, 647, 813

\bibitem[Oxley\ \etal(2004)]{oxley_etal_2004}
Oxley, P. \etal, 2004, in Barnes, W.L. \&\ Butler, J.J., eds,
Proc. of the SPIE, Earth Observing Systems IX, 5543, 320

\bibitem[Page\ \etal(2007)]{page_etal_2007}
Page, L. \etal, 2007, \apjs, 170, 335

\bibitem[Ponthieu\ \etal(2005)]{ponthieu_etal_2005}
Ponthieu, N. \etal, 2005, \aa, 444, 327 

\bibitem[Prunet\ \etal(1998)]{prunet_etal_1998}
Prunet, S., Sethi, S. K., Bouchet, F. R., Miville-Desch\^enes, M.-A.,
1998, \aa, 339, 187 

\bibitem[Reichardt\ \etal(2008)]{reichardt_etal_2008}
Reichardt, C.L. \etal, 2008, submitted to \apj

\bibitem[Shewchuk(1994)]{shewchuk1994}
Shewchuk, J.R., 1994, preprint, {\tt http://www.cs.cmu.edu/$\sim$quake-papers/painless-conjugate-gradient.ps}

\bibitem[Schlegel\ \etal(1998)]{schlegel_etal_1998}
Schlegel, D.J., Finkbeiner, D.P., \& Davis, M.,  1998 \apj, 500, 525

\bibitem[Spergel\ \etal(2007)]{spergel_etal_2007}
Spergel, D.N. \etal, 2007, \apjs, 170, 377 

\bibitem[Stolyarov\ \etal(2005)]{stolyarov_etal_2005}
Stolyarov, V., Hobson, M.P., Lasenby, A.N., \&\ Barreiro, R.B., 2005, \mnras, 357, 145

\bibitem[Stompor\ \etal(2002)]{stompor_etal_2002}
Stompor, R. \etal, 2002, \prd, 65, 022003

\bibitem[Zaldarriaga and Seljak(1997)]{zaldarriaga_seljak_1997}
Zaldarriaga, M. and Seljak, U., 1997, \prd, 55, 1830

\bibitem[Zaldarriaga and Seljak(1998)]{zaldarriaga_seljak_1998}
Zaldarriaga, M. and Seljak, U., 1998, \prd, 58, 023003

\bibitem[Zaldarriaga(2001)]{zaldarriaga_2001}
Zaldarriaga, M., 2001, \prd, 64, 103001

\end{thebibliography}
\end{document}